\documentstyle[12pt]{article}          
\def	\eqnum		#1{(\ref{#1})}       
	\newdimen\eqskip
	\newdimen\txtskip
\def	\be		{\begin{equation}}
\def	\ee		{\end{equation}}
\def	\ba		{\begin{eqnarray}}
\def	\ea		{\end{eqnarray}}
\def	\nn		{\nonumber}
\def	\=		{\;=\;}
\def	\frac		#1#2{{#1 \over #2}}

\def	\to		{\rightarrow }

\def	\eg		{e.g.}
\def	\as		{\mbox{$\alpha_{\rm s}$}}
\def	\orderas	#1{${\cal O}(\as^#1)$}
\def	\oas		{${\cal O}(\as)$}

\def	\muf		{\mbox{$\mu_{\rm F}$}}
\def	\mur		{\mbox{$\mu_{\rm R}$}}

\def	\z		{\mbox{$Z$}}
\def	\w		{\mbox{$W$}}

\def	\b		{\mbox{$b$}}

\def    \pt	        {\mbox{$p_t$}}

\def    \missEt         {\ifmmode{/\mkern-11mu E_t}\else{${/\mkern-11mu
E_t}$}\fi}

\def 	\mb   		{\mbox{$m_b$}}

\def 	\mq   		{\mbox{$m_Q$}}
\def 	\mtr  		{\mbox{$m_T$}}

\def 	\lqcd  		{\mbox{$\Lambda_{QCD}$}}
\def 	\lff   		{\mbox{$\Lambda^{(2)}_{4}$}}
\def	\jpsi		{\mbox{$J/\psi$}}
\def	\psitwos	{\mbox{$\psi(2S)$}}

\begin{document}
\begin{titlepage}
\begin{flushright}
IFUP-TH 4/93 \\
FERMILAB-PUB-93/19-E\\
hep-ph/9302306
\end{flushright}
\vfill
\begin{center}
       {\Large  \bf \sc QCD Tests in Proton-Antiproton Collisions} \\[0.35cm]
	(To appear in Annu. Rev. Nucl. Part. Sci., Vol. 43) \\[0.7cm]
\it John E. HUTH
       \footnote{Fermi National Accelerator Laboratory, Batavia IL, USA.
Bitnet:
       	HUTH@FNALD.FNAL.GOV}
       \footnote{Address after Sept 1$^{\rm st}$, 1993: Dept. of Physics, Lyman
       Laboratory,  Harvard University, Cambridge, MA 02138, USA. Bitnet:
       HUTH@HUHEPL.HARVARD.EDU}
and
Michelangelo L. MANGANO
   \footnote{Istituto Nazionale di Fisica Nucleare, Scuola Normale Superiore
   and Dipartimento di Fisica dell'Univerita', Pisa, ITALY. Bitnet:
	       	MLM@FNALD.FNAL.GOV} \\[0.8cm]
\end{center}
\vfill
\tableofcontents
\vfill
\end{titlepage}
\section{INTRODUCTION}
Quantum Chromodynamics (QCD) \cite{qcd1,qcd2}
is regarded as the best, if not the only,
viable theory of the  strong interactions.  Recent theoretical and
experimental developments have significantly increased our ability to
perform quantitative tests \cite{altarev} and have deepened our
understanding of hadronic interactions.   This is particularly
true for higher energy processes where the decreasing value of the
coupling constant $\alpha_s(\mu)$ allows reliable results from
a perturbative expansion \cite{DDT}-\cite{physrep}.  In addition,
the non-perturbative transition from the fundamental objects in the theory
(quarks and gluons) to observed particles (hadrons) has a smaller influence
on measured quantities. At the energies now
accessible, one is expected to be far
away from the long distance regime where  perturbation theory breaks down, and
from the ultra-short distance regime where one might witness the onset of
new dynamics. This,
however, does not excuse one from vigilance for significant deviations from
perturbative predictions.

Tests of QCD in hadron-hadron collisions display a parallel
development in both theory and experiment.  The earliest measurements
of high $p_t$ hadron production at the ISR belied the then hidden partonic
component of the proton.
The observation of jets and early tests
of QCD at the S$\overline p p$S \cite{dilella} were largely
qualitative, yet
they demonstrated the predictive power of the theory at leading
order in perturbation theory.  Currently we are entering a period
where the emphasis is being placed on measurement
precision.  From an experimental standpoint this implies making
measurements with high statistics and small systematic uncertainties.
{}From a theoretical standpoint,  this means calculating quantities
at successively higher orders in perturbation theory, and using
constraints from a number of sources (c.f. parton distribution
functions) to pin down predictions.   In the interplay between
theory and experiment, there must be a coherent view of how
quantities are defined (e.g. what precisely is a ``jet''?) in
order to arrive at definitive tests.

Recent studies with high statistics of hadronic decays of
the $Z^0$ from $e^+e^-$ production have yielded impressive
new confirmations of the theory \cite{leprev}.
QCD tests in $p \overline p$ collisions are not as direct as those in
$e^+e^-$ owing partly to the complications associated with partons
in the initial state and the beam fragments (the so-called ``underlying
event'').  Accompanying this complexity is however a richness which allows
one to attack a given problem in a number of complementary ways.  For example,
knowledge of  the partonic density as a function of the proton momentum
fraction introduces uncertainties in the predictive power of the  theory.  On
the other hand the same feature allows one to obtain data which span a  wide
range of center-of-mass energies in the parton-parton  frame for a fixed set
of beam conditions. The variety and diversity of hard processes accessible  in
hadronic collisions, together with the enormous cross sections and energy
reach, provide us with a multitude of phenomena  inaccessible to current
$e^+e^-$ experiments \cite{ehlq}-\cite{barger}.

A major ingredient for the prediction of cross sections in $p \overline p$
collisions is the distribution of partons inside the proton
\cite{mishra,tungrev}.
Recently there has been significant progress in theory and
experiment, leading to expanded measurements of parton distribution
functions (PDF).  Concurrently, next-to-leading order (NLO) perturbative
calculations have been performed for most interesting processes.
As a result we have collected a substantial body of evidence
demonstrating that QCD properly
describes this physics both qualitatitively and quantitatively.
However, there are some outstanding questions which need to
be resolved.
The aim of this review is to present this evidence in a critical way,
pointing  out where theoretical and experimental improvements are
expected or desired and where one can rely on QCD to extract new
information.

We review the last analyses from the CERN Collider experiments (UA1 and UA2)
which have completed operations (1991) along with data from     the Collider
Detector at Fermilab (CDF) dating mostly from the 1988-89 running period of the
Fermilab Tevatron. A new cycle of data taking has started in the Summer 1992 at
the Fermilab Collider, with the presence of a new experiment, D0. Several of
the results presented here are still on their way to press  and only available
in preprint form. We felt it necessary to include them because they often add
important new contributions to the overall picture. We regret that the analyses
from the latest set of data collected since the Summer 1992 are still premature
to appear in this review, and we look forward to their completion for the
important implications they will have on the study of QCD in hadronic
collisions.

\section{QCD IN HADRONIC COLLISIONS}
One of the fundamental properties of QCD is the shrinking of the coupling
constant as the energy of the interaction grows (asymptotic
freedom). This implies that perturbative techniques can be used to
study high energy hadronic phenomena. In spite of this, we cannot fully rely on
perturbation theory (PT) because the fundamental particles whose interactions
become weak at high energy are deeply bound inside the hadrons we use as beams,
targets, or as observables.
The solution is given by factorization theorems
\cite{collins,PQCD}, whereby cross sections can be expressed as the product of
factors, each one involving phenomena appearing at different energy scales.

In the case of hadronic collisions, the separation of the initial state
evolution
from the hard perturbative interaction can be
represented, for a generic inclusive process $A+B\to C+X$, as:
\be \label{efact}
	\sigma_{A+B\to C+X} \= \nn \\
  	\sum_{ij}\int {\rm d}x_1 \, {\rm d}x_2 \;
	f_i^A(x_1,\muf) \, f_j^B(x_2,\muf)
	\hat\sigma_{ij\to C}(\hat s,\muf,\mur,\as(\mur)) \; ,
\ee
with $\hat s=x_1x_2s$. $i$ and $j$ are indices for any pair of partons (quarks
or gluons) contributing to the process, $f_i^A(x,\muf)$ represents the number
density of partons of type $i$ carrying a fraction $x$ of the momentum of the
parent hadron $A$, and $\hat\sigma$ is the cross section for the elementary
hard process,
calculable in perturbation theory \footnote{If $C$ were a specific hadron,
an independent factorization theorem for fragmentation would apply.
$\hat\sigma$ will then be the convolution of a purely partonic process with a
fragmentation function describing the transition of a final state parton into
$C$.}.
Furthermore, the parton distribution functions (PDF) $f(x,\muf)$ are
independent of the specific reaction.
The universality of PDFs is a key property, since they are not calculated from
first principles as they contain non-perturbative information.
They can then be extracted from one process and applied to predict rates for
another one.

The scale \mur\ introduced above is the scale at which the ultraviolet
singularities of the theory are subtracted (``renormalization prescription''),
determining the ``running'' of \as(\mur). The energy scale \muf\ represents
the freedom given by the factorization theorem to absorb as much or as little
of the radiation from the evolution of the initial
state parton into the PDF, including the rest in $\hat\sigma$ (``factorization
prescription'').

The final result should not depend on the choice of \muf\ and
\mur. This is the case if we evaluate  $\hat\sigma$, \as(\mur)
and $f(x,\muf)$ exactly. Any fixed order perturbative approximation will leave
a residual dependence on \muf\ and \mur.  This dependence is logarithmic, and
the sensitivity of a fixed-order cross section to variations of $\mu_{\rm F,R}$
is usually taken as an estimate of the importance of neglected higher order
terms. Since the two scales have different origins, they do not have to be
the same. Nevertheless it is customary to take them equal and of the order of
the energy scale of the hard subprocess, to avoid the appearance of
logarithms of large ratios in the perturbative expansion and minimizing the
effect of higher order terms.
The invariance of the results under changes of \muf\ allows to formulate
an equation (the Altarelli-Parisi (AP) equation,
\cite{ap}), which ``evolves'' the PDF from  one scale \muf\ to another.
With this equation, measurements of the PDF carried out at relatively low
values of $\mu$ in Deep Inelastic Scattering (DIS) experiments can be used to
extrapolate the values of parton densities  to the values of $\mu$ found in
hard hadronic collisions.

The cross sections derived at leading order (LO) have large  uncertainties
associated with the choice of $\mu$, since the matrix elements at this
order do not contain any initial state radiative process, and are thus
independent of \muf.  A dependence on \muf\ appears inside $\hat\sigma$ only at
the next order in PT (via the subtraction of the
initial state collinear singularities) and a partial cancellation  between
$f(x,\muf)$ and $\hat\sigma(\muf)$ takes place.  It is therefore  important to
have available at least the NLO matrix elements to carry out quantitative tests
of QCD. In spite of the technical difficulties, the calculation of most of the
interesting processes has been completed today at NLO accuracy, and new
techniques are being developed to enable the calculation of yet higher order
corrections \cite{physrep,giele,bernnlo}.

Likewise, analyses of the PDFs have been carried out in recent years with
similar precision \cite{dflm}-\cite{CTEQ},  providing the necessary elements
for consistent NLO calculations. We refer to Ref.~\cite{mishra,tungrev}\ for a
review of PDF  measurements and parameterizations, and limit ourselves to
point out the existence of recent data extending the measurements of
$F_{2,3}(x,Q)$ down to $x=0.008$ for $Q^2$ as large as 5 GeV$^2$
\cite{nmc,CCFR}. These data show clear discrepancies with previous
extrapolations of $F_2$ to small-$x$, indicating a violation of the light
flavour symmetry in the sea densities. Nevertheless the measurements
confirm \cite{CCFR}\  earlier estimates of the behaviour of the gluon density,
whose extrapolation to small-$x$ is responsible for systematic  uncertainties
in the calculation of most hadronic processes.
New fits to these data have appeared \cite{newmrs,CTEQ}\
and show that the gluon density is now
under a rather solid control in the region  $x>0.01$ and $Q>10$ GeV
(Figure~\ref{fnewpdf}). This is the region of sensitivity for most QCD
processes probed by current hadron colliders.

The formalism described so far only allows the calculation of inclusive
quantities. This is the case of jet distributions and correlations, or
electroweak boson cross sections. The inclusive nature of the PDFs by itself
prevents predictions on the structure of the radiation emitted during the
initial state evolution. A more exclusive picture of the event structure is
often required, both as a tool to understand the experimental systematics
(calorimeter energy response, effect of particle isolation requirements, etc.),
and as a way of probing more specific predictions of QCD (\eg\ jet
fragmentation properties).

To complete the description of the event structure, a complementary approach,
known as shower Monte Carlo (MC), has been  developed~\cite{webberrev}. Here
the partons from a hard collision evolve via gluon and quark radiation, until a
small virtuality scale $Q_0$ is reached where $\as(Q_0)$ is large. Here
confinement forces take over, hadronizing the colored partons. Descriptions of
the hadron-formation phase can be included  \cite{string,webber}\ and tuned
using a reference process (e.g. jet production in $e^+e^-$ collisions). These
non-perturbative effects are believed to be  universal, namely they do not
depend on the hard process. The main features of the final state of
different processes are thus accounted for by the QCD evolution, as the
distributions of the hadrons are expected to closely mimic those of
the partons they originated from (Local Parton-Hadron Duality, LPHD
\cite{LPHD,amati}).

Such models have been implemented in several computer programs
\cite{odorico}-\cite{herwig}. They differ from one another in several aspects,
ranging from the accuracy of the  perturbative evolution, to the hadronization
scheme. The spectrum of the radiation is given by perturbative QCD, and, in
some cases \cite{herwig}, it includes all orders of leading and
large classes of subleading soft and collinear logarithms \cite{BCM,CMW}.
MCs such as PYTHIA \cite{pythia}\ and HERWIG \cite{herwig}, finally, succeed in
describing typical quantum mechanical effects due to color interference via a
simple ``angular ordering'' prescription, which limits the kinematical phase
space available for the emission of soft gluons from colored currents
\cite{webberrev,LPHD}.

\section{JET PRODUCTION}
\subsection{Inclusive Jet Production}

The precision of QCD tests involving jets has been limited by the necessary
correspondence between the final state  sprays of hadrons and the partons from
a hard  scattering, whose cross sections are perturbatively calculable. There
have been substantial developments resulting from
the higher center-of-mass energies and an improved understanding of
experimental systematics.
The recent calculation of jet cross sections beyond LO
in PT has reduced the theoretical uncertainties
greatly and predicts new quantities.

The inclusive jet cross section,  $\sigma(p \overline p \rightarrow$ JET$+X)$
is the most straightforward quantity to test.
At LO (${\cal O}(\alpha_s^2)$),  eight
diagrams contribute to the scattering and give rise to two parton final states
\cite{localc}.
At ${\cal O}(\alpha_s^2)$, the partonic cross section is directly equated to
the measured jet cross sections.
For a fixed $p \overline p$ center-of-mass energy, the
inclusive cross section is a non-trivial function of two variables: $\eta$,
the jet pseudorapidity ($\equiv \log\,\cot\, \theta/2$, where $\theta$ is the
polar angle), and
the transverse energy, $E_t$.
As will be discussed later, the issue of how precisely one defines
jet $E_t$ is important to the overall consistency of the comparison
between theory and experiment.  For now it can be taken to be the
sum of the transverse energies of discrete sub-units, be it particles
or calorimeter towers.
The most common representation of
the data is typically in terms of the differential cross section,
$d \sigma / d E_t$; this is really an average of the inclusive
cross section over some pseudorapidity interval in a detector:
\begin{equation}
<{d \sigma \over d E_t} > \equiv {1 \over \Delta \eta}
\int_{-\eta}^{+\eta} {d \sigma \over d E_t d \eta} d \eta
\end{equation}
In some cases this is expressed as the cross section evaluated
at $\eta = 0$: ${ d \sigma \over d E_t } |_{\eta=0}$, assuming that
the rate is constant in $\eta$ over a large enough interval.  Most
collider experiments report measurements in  roughly the central two units of
pseudorapidity.

As mentioned in Section~2, large uncertainties are associated in LO to
changes in the factorization/renormalization scale $\mu$.
For a range of $2E_t > \mu > E_t/2$, the LO
cross sections for $d \sigma / d E_t$ vary by approximately 50~\%.    This
uncertainty is roughly a constant multiplier of the cross section for different
$E_t$'s , with only a modest dependence of the shape of the cross section as a
function of jet $E_t$.

Despite the large uncertainties,
if one chooses a renormalization scale
($\mu = E_t/2$), and compares data to QCD for Tevatron
($\sqrt{s}=1.8$ TeV)\cite{cdfjets}, S$p \overline p$S ($\sqrt{s}=630$ GeV)
\cite{ua2jets,ua1jets} and ISR ($\sqrt{s}=63$ GeV) \cite{isrjets},
one finds an impressive agreement
between the experimental results and the theory with only one
floating parameter.  This is shown in Figure \ref{locompar}.

The UA2 collaboration also measured the jet cross section
for different pseudorapidity intervals \cite{ua2jets}.
Although not shown here, the agreement is good in the
central region ($|\eta| < 0.8$), but for larger
values of pseudorapidity ($1.2 < |\eta| < 2.0$) is marginal \cite{ua2jets}.
There is no clear explanation for this \cite{ua2jets}.

To calculate the cross section at
NLO, one must combine
graphs where a parton is radiated and
loop diagrams (Figure \ref{nlodiag}).
At this order, factors of $\log(\mu)$ appear which cancel
some of the $\mu$ dependence in $\alpha_s(\mu)$ and the PDFs.
The evaluation
of the full NLO matrix elements was initiated in Ref.~\cite{oldnlo}\ and later
completed by Ellis and Sexton \cite{sexton} in
1986. A confirmation of these results using a different approach has come
recently from Ref.~\cite{bernnlo}. These works did not include an
explicit calculation of the cross
section.
Whereas at LO a direct correspondence is made between
jet and partonic cross sections, the situation is not as
straightforward at NLO.
In order to evaluate the cross section at ${\cal O} (\alpha_s^3)$
(in fact even to obtain finite results \cite{furman}), one
must specify what a jet is at the partonic level.  If two partons are
close together they may be merged into a single ``jet''.  Here one
speaks only of jet, as opposed to partonic cross sections at both the
theoretical as well as experimental levels.  Ideally, the theoretical
jet definition should thus
be as close as possible to the experimental jet definition.

Aversa, Chiappetta, Greco and Guillet
\cite{aversa}\ and Ellis, Kunszt  and Soper \cite{kunzst}\ have
used the matrix elements of Ref.~\cite{sexton}\
to derive jet cross sections at ${\cal O}(\alpha_s^3)$.
Although both groups employ different computational techniques, the
results have been shown to be numerically identical \cite{mjjpaper}.
After the imposition of a jet definition (see below),
the NLO cross sections show substantially
smaller sensitivities to renormalization scale variations than at
LO. Over a range of renormalization
scales close to the hard scattering scale ($E_t/4 < \mu < E_t$),
the uncertainties in the cross section have been reduced from
50~\% to 10~\% over most of the range of accessible $E_t$.
The inference is that the effects of still higher
order contributions are rather small at ${\cal O}(\alpha_s^3)$.
Figure \ref{renorm} shows the variation of the ${\cal O}(\alpha_s^3)$
and ${\cal O}(\alpha_s^2)$ cross sections with $\mu$ for
100 GeV jets.  At LO,
one finds a large monotonic variation of the cross section with
$\mu$, whereas at ${\cal O}(\alpha_s^3)$, the negative $\log(\mu)$
contributions  from the virtual terms reduce the cross section at very
small $\mu$.
Note that the sensitivity of the cross section at
renormalization scales near the hard scattering scale is greatly
reduced at NLO.

Several experimental jet algorithms have been employed.  When
cross sections are derived only at LO and when
uncertainties are large, these differences can be forgiven; even so,
comparisons between experiments are rendered difficult.  For
example, UA1 \cite{ua1alg} and CDF \cite{3jetprd}
employed cone algorithms, whereas UA2
used initially a nearest neighbor algorithm \cite{ua2alg}.
A typical hadron collider algorithm is the ``cone'' algorithm, which
has been suggested as a standard for $p \overline p$
experiments \cite{huth_snow}.
It operates in a space defined by pseudorapidity
and azimuth ($\eta - \phi$) on particles, or partons or calorimeter
towers, depending on the specific application.
In this metric, one can define a jet to be the partons or
particles found
in cones or, more precisely, circles of radius $\Delta R \equiv
\sqrt{\Delta \phi^2 + \Delta \eta^2}$.  The transverse energy,
$E_t$, is the sum of the transverse energies of particles, partons or
calorimeter towers inside a fixed radius.   The direction of a jet
in $\eta$ and $\phi$ can be defined as the $E_t$ weighted centroids:
\ba
E_{t}^{jet} &=& \sum_i E_{t}^i \\
\phi_{jet} &=& (\sum_i E_{ti} \phi_i )/ E_{t}^{jet} \\
\eta_{jet} &=& ( \sum_i E_{ti} \eta_i ) / E_{t}^{jet}
\ea

The above description is not complete,
however.   It does not tell where to initially place the cones
to form the above quantities, and does not describe how to handle
cases where cones overlap (``merging'').
In the case of experiments employing calorimeters, the
initial jet direction can be defined by towers with $E_t$ above
a given threshold (seed towers).
An iterative approach can be adopted
to find a stable center of the cluster by successively
recomputing the cluster centroid until the list of towers or
particles in the cone is stable \cite{3jetprd}.
If two jets
are greater than one cone radius apart, but
less than two radii ({\it i.e.} $R_o < \Delta R_{1,2} < 2R_o$)
should they be identified as one or two jets?   The inherent
difference between two partons in a calculation and calorimeter
towers in an experiment can make it difficult to
achieve a precise uniformity in the jet definition.

Both CDF and the UA2 experiments have measured jet cross sections
using cone algorithms with $R=0.7$ and $1.3$ respectively,
with reduced uncertainties \cite{cdfjets},\cite{ua2jets}.
Although the UA2 results are not shown, there is good agreement with the
${\cal O} (\alpha_s^3)$ predictions despite the fact that the
calculations do not apply strictly for $R > \pi/3$ \cite{kunzst}.
The dominant
experimental uncertainties are associated with the hadronic
energy scale.  The calorimeter response to jets, particularly the
hadronic component, is difficult to calibrate in an absolute way.
There are no test beams with monoenergetic sources of jets, so
the calorimeter response must be derived from a convolution of the
calorimeter response to hadrons of varying energy (including
$\pi^o$'s) with the jet fragmentation spectrum.  Although the
response can be checked with sources such as jets recoiling against
direct photons, there is no unimpeachable source on which
to calibrate.   Typical energy scale uncertainties are now
 $\approx$4-15\%  in $\delta E / E$.  A systematic shift in energy
scale is equivalent to an uncertainty in the cross section.
Since the cross section is typically a steeply falling function
of $E_t$, following a power law spectrum of $E_t^{-5}$,
the resulting uncertainty in jet cross section is 20-75\%.
Recent work by both the UA2 \cite{ua2jets} and CDF \cite{cdfjets}
collaborations pressed the lower bounds of these uncertainties,
thus improving the level of comparison to theory.  The uncertainty
can be expressed as an overall multiplicative factor which is
independent of jet $E_t$ (20\% and 35\% for CDF and UA2 respectively)
and a smaller term
which is $E_t$ dependent and can be roughly 5\% \cite{cdfjets}.

The agreement appears to be very good on a logarithmic scale.
To illustrate significant features of the comparison, however, one can
plot the cross section on a linear scale, as a ratio of
(Data -- Theory)/Theory as a function of jet $E_t$.  Figure
\ref{linear} shows such a comparison for CDF data \cite{cdfjets}.
The QCD ${\cal O}(\alpha_s^3)$ prediction for $\mu = E_t/2$ is
defined to be the ``Theory'' or 0 on this plot for the purposes
of normalization.  The data have uncertainties factored into a
combination of the $E_t$ dependent systematic and statistical
uncertainties which are displayed on the error bars, and an
$E_t$ independent component which is 20~\%.   Figure \ref{linear} (a)
illustrates the improvement in the uncertainty associated with
theory for a variation of $E_t/4 < \mu < E_t$.  One can see that
the uncertainties are substantially reduced at ${\cal O}(\alpha_s^3)$.
Figure \ref{linear} (b) shows the effect of different PDFs
\cite{hmrs,mt}
on the predicted cross section.    As one can see there is
some dependence on the shape of the derived cross section on the choice
of PDF, however, the overall agreement is quite good.

This does not exhaust comparisons at ${\cal O}(\alpha_s^3)$.
Figure \ref{conevar} shows the variation of the cross section with
cone size for $E_t=$100 GeV jets from CDF compared
with ${\cal O}(\alpha_s^3)$ predictions \cite{wainer,eksprl,aversa2}.
The data display the statistical errors only, but the $\approx $
23~\%
systematic uncertainties are largely independent of $R$.
Since there are only two back-to-back partons in the
LO calculation, one can only predict such a variation
beginning at NLO.
An interesting feature of the calculation is the minimal
sensitivity to $\mu$ for a cone size of $R=0.7$, whereas the
sensitivity is much greater both for $R< 0.5$ and $R>0.9$.
{}From this standpoint
$R=0.7$ represents an ``optimal'' cone
size for comparison to ${\cal O}(\alpha_s^3)$ predictions.
The data appear to be in rough agreement with at least one
of the QCD predictions ($\mu = E_t/4$), but on the whole, there
seems to be a trend for the data to show a slightly steeper
dependence on $R$ than the theory predicts.

A quantity related to the variation of cross section with
cone size is the jet profile.
To measure this, one
can pick a large radius ($R=1.0$), and then examine the fraction of
the jet $E_t$ contained in a smaller sub-cone of radius $r$:
$F(r,R,E_t)$.
CDF measured this quantity using charged particle tracking data
because it is more fine-grained than calorimetric information.
Figure \ref{flow} shows a plot of $F(r,R,E_t)$ from CDF data.  Also
shown are the predictions of ${\cal O}(\alpha_s^3)$ QCD \cite{eksprl}
for different
choices of renormalization scale.   It is perhaps surprising that the
data are so well described at the level of just one gluon bremsstrahlung
when there are typically 10 charged hadrons in a typical jet.
Since ${\cal O}(\alpha_s^3)$ is the lowest order at which one can
speak of a jet profile, the sensitivity to renormalization scale
is fairly large.

There is an apparent contradiction between the profile measurement
and the variation of the cross section with cone size.
One naively might expect that if there were good agreement between
data and theory for one quantity, having chosen a renormalization
scale, that there would be a good agreement for the other.  This
expectation is based on the assumption that the variation of cross
section with cone size just depends on the energy flow within the
cone.
This assumption is not valid, however.  In the inclusive measurement,
jets are clustered independently for each cone size, $R$, chosen,
whereas for the jet profile, only a single cone of $R=1.0$ is
used.  The main difference is the ``merging'' step.  Ellis,
Kunszt and Soper \cite{eksprl} have examined the effect of merging
in the ${\cal O}(\alpha_s^3)$ predictions. As discussed above,
there is an ill defined region where two partons may be
separated by a distance $R_o < \Delta R < 2R_o$.  In order to
mimic the merging in the experimental algorithm, partons are merged
into a single jet if they have a separation $\Delta R < R_{sep}$.
The calculation implicitly had $R_{sep}=2.0$.  However as seen in
Figures \ref{conevar} and \ref{flow} a value of $R_{sep} = 1.3$
and choice of $\mu = E_t / 4$ fit both distributions \cite{eksprl}.
Although one has obtained consistent results, one has done so at the expense
of adding a tunable parameter to the theory.

The ambiguities related to the prescription for the merging of jets are absent
in the class of jet definitions generally used by the $e^+e^-$ experiments. The
prototype of these jet definitions is provided by  the JADE algorithm
\cite{jadealg}, which builds clusters of charged particles  according to an
invariant mass cut. The invariant mass normalized by the center-of-mass energy,
$y_{ij}=M_{ij}^2 / E_{cm}^2$, is used to define jets as distinct objects.
$M_{ij}$ is the invariant mass of pairs of particles or of a particle and a
cluster. At each step of an iterative procedure, the pair with the smallest
$y_{ij}$ is merged into a new cluster if $y_{ij}<y_{cut}$. If no pair is left
passing the cut, all remaining clusters are called jets. The leading weakness
of the JADE algorithm from the point of view of $p p$, $p \overline p$ and $e
p$ colliders is that all particles are associated to some jet, including those
coming from the underlying event and which do not belong to the hard process.

Improved versions of the JADE algorithm have recently been proposed
\cite{durham}, which reduce the  sensitivity to the jet definition under
hadronization corrections,  and make it possible to resum large classes of
leading and subleading perturbative corrections in the theoretical
calculations. These prescriptions can be extended \cite{durhamDIS}\ to
processes with hadronic  the initial states. In this formulation they provide
an unambiguous prescription for the merging of jets and allow the universal
factorization of initial state collinear singularities, minimizing the
contamination from the hadron remnants and the underlying event. The similarity
with the $e^+e^-$ jet definitions will make it possible to compare jet
properties between $e^+e^-$ and hadron colliders in a consistent and universal
fashion. No complete phenomenological study of this new algorithm is available
as yet, but we hope that progress will be made soon (S~Ellis, Z~Kunszt \&
D~Soper, personal communication) and that experimental measurements will follow
as well.

\subsection{$x_t$ : Jet Scaling with ${s}$}
If one plots the inclusive jet cross section in terms of two
dimensionless variables, the ``scaling'' hypothesis predicts
an independence of $p \overline p$ center-of-mass energy, ${s}$.
In reality, the evolution of PDFs and $\alpha_s$ with the hard scattering
energy scale
causes a violation of scaling for the inclusive jet cross section.

To test scaling, one typically plots $E_t^4$ times the invariant
cross section ($E \, d\sigma^3/dp^3$) as function of
$x_t \equiv  2 E_t /\sqrt{s}$ to obtain two dimensionless
quantities to express the jet cross section.  If scaling were
valid, cross sections measured in this way at any $\sqrt{s}$ would
all fall on a single universal curve.  QCD, on the other hand,
lifts the degeneracy.
The predicted ratio of cross sections at two different
center-of-mass energies as a function of $x_t$
is relatively insensitive to choice of PDF,
renormalization scale or the order of the calculation,
making it a relatively solid test of the theory.
Independent measurements made at the
S$p \overline p$S, ISR and Tevatron showed rough agreement with
QCD scale breaking \cite{ua2jets}.
CDF \cite{behrends}
have recently compared jet cross sections at $\sqrt{s}$ = 546 and
1800 GeV as a test of $x_t$.  When the cross sections are measured
in one experiment, a large part of systematic  uncertainties
(e.g. hadronic energy scale) cancel when
the ratio of the cross sections is taken, improving substantially
the level of comparison.

Figure \ref{xt} shows the ratio of scaled cross sections as
a function of $x_t$ for CDF data taken at both center-of-mass
energies.  The error bars show statistical uncertainties, and
the shaded area  indicates an overall systematic uncertainty in the
ratio.  The data are clearly inconsistent with scaling
(Ratio=1).
The data do not exhibit a wonderful agreement with
QCD either.

The discrepancy with QCD is about 2 standard deviations in the
systematic uncertainty, which is not sufficient to indict QCD
by any means, but is curious.   At the moment, there is no
obvious explanation for such a discrepancy.
PDFs in the relevant $x$ range  ($x > 0.1$) have
been measured by a number of DIS.
Further running of the Tevatron collider at a lower
center-of-mass energy could shed light on this.

\subsection{Two-jet Distributions}

The ${\cal O}(\alpha_s^3)$ predictions have been extended to measurements
where one defines a two-jet inclusive final state.
Since it is impossible to either measure or calculate states beyond
LO with
two and only two jets, due to soft radiation, one can form quantities
from the leading two jets, and ignore other energy in the event.
Recently, Ellis,
Kunszt and Soper \cite{mjjpaper} have extended the ${\cal O}(\alpha_s^3)$
calculations to predict
the two jet invariant mass and center-of-mass angular distributions.
Such distributions are sensitive to the presence of deviations from
QCD arising from quark compositeness, technicolor \cite{lane} and
axigluons \cite{axigluons}.

The CDF two-jet invariant mass distribution $M_{jj}$ is defined as:
\begin{equation}
M_{jj} \equiv \sqrt{(E_1 + E_2)^2 - ({\bf p_1 + p_2)^2}}
\end{equation}
where $E_i$ and ${\bf p_i}$ are the energies and momenta of
jets $1$ and $2$.  Note that the effective  masses of
the jets enter into the determination of $M_{jj}$.  The jet
mass, an internal quantity, can be associated with gluon
bremsstrahlung within the clustering cone.
The CDF $M_{jj}$
cross sections were determined for cone sizes of 0.7 and 1.0
\cite{orso}.
For the HMRSB \cite{hmrs} and MT S1 \cite{mt}\ PDFs
the ${\cal O}(\alpha_s^3)$ predictions
appear to be in good agreement with the data for a clustering
cone of 1.0, but disagree at about the 2 standard deviation
level for a cone of 0.7 \cite{orso}.

The dijet angular distribution has likewise been calculated at
${\cal O}(\alpha_s^3)$ \cite{mjjpaper}. Since
invariant mass and $\cos(\theta^*)$ are independent variables,
the data can be placed in different bins of $M_{jj}$.  Here
$\theta^*$ is the center-of-mass polar scattering angle.
Since the cross section is dominated by $t$ channel
exchange, it rises very rapidly with increasing $cos(\theta^*)$ and
it is more convenient to plot the data as a function of the
variable $\chi$, defined as:
\begin{equation}
\chi \equiv {1 + cos \, \theta^* \over 1 - cos \, \theta^*}
\end{equation}
If plotted versus $\chi$, the Rutherford scattering pole is
taken out.   There is a rise in cross section for $\chi \approx 1$
(90 degrees) associated with the contribution of $s$-channel
scattering. Figure \ref{angular} shows the results of an analysis of $dN/d\chi$
by CDF using a cone size of $0.7$ \cite{mueller}.
The data are divided into three bins
of $M_{jj}$. One can see that the data are well described by
both ${\cal O}(\alpha_s^2)$ and ${\cal O}(\alpha_s^3)$ QCD predictions.
The data are separately normalized for each bin of $M_{jj}$.

The effects of quark compositeness would be to increase the amount
of data found near $\chi \approx 1$.  Quark compositeness is typically
parameterized in terms of a four-fermion interaction with a coupling
inversely proportional to a characteristic energy scale
(related to the ``size'' of the quark), $\Lambda_c$ \cite{eichten,cdfjets}.
Such an interaction gives rise to an
isotropic distribution in the center-of-mass system, and
also contributes a rising cross section at large $E_t$ or $M_{jj}$.
In order to search for compositeness, one can take several bins
of $M_{jj}$ and examine the dijet angular distribution in each.
Compositeness could be manifest as
an increase in 90 degree scattering in the highest $M_{jj}$ region,
while the remaining data should be well described by QCD.
The CDF data have allowed limits
to be placed on $\Lambda_c > 1.4$ TeV using the inclusive jet data
\cite{cdfjets},
and at $\Lambda_c > 1.0$ TeV using the angular distribution \cite{mueller}.

\subsection{Jet Fragmentation}
As discussed in Section~2,
some aspects of jet fragmentation reveal the underlying QCD mechanisms,
particularly when one assumes that the behavior of hadrons in jets mimics
features of the partonic emission. On the basis of this, one expects for
example that gluon jets will have softer fragmentation than quark jets and that
average multiplicities will increase with energy.

Studies of jet fragmentation in hadronic collisions have been performed in the
past by UA1 and UA2\cite{ua1frag},  providing the first indications that jets
in $\bar p p$ reactions have higher multiplicities than in $e^+e^-$
annihilation.

The most notable quantity to study
is the jet fragmentation function, which describes the probability of
finding a hadron carrying a given fraction of the jet's momentum.  This
is typically described in terms of the component of hadronic momentum
parallel to the jet axis: $p_{\parallel}$.  The
charged particle fragmentation function,
$F(z)$ , is defined as ($z \equiv p_{\parallel} / p_{jet}$):
\begin{equation}
F(z) = {1 \over N_{jet}} {dN_{ch} \over dz}
\end{equation}

The evolution  of $F(z)$ as a function of the hardness of the primary collision
is a good test of QCD, and, in principle, can be used to extract $\alpha_s$.
The same mechanism for the evolution of parton densities, namely
soft and collinear parton emission, is responsible for the logarithmic
evolution of $F(z)$.
Figure \ref{evol} shows the evolution of different bins of $F(z)$ as a function
of dijet invariant mass ($M_{jj}$)  from CDF data \cite{hubbard}. The data
agree well with a logarithmic evolution with $M_{jj}$ and have a distinct
similarity with data from $e^+e^-$ \cite{tasso}, which are plotted as a
function of $\sqrt{s}$.   $M_{jj}$ appears to be  a sensible variable to
express this evolution insomuch as it  is a measure of the hardness of the
scattering, particularly in the central pseudorapidity region.

Notice however that the $e^+e^-$ and $\bar p p$ curves do not match:
the $e^+e^-$ curve corresponding to the lowest $z$ bin extrapolates below the
equivalent CDF curve. This behaviour is consistent with the notion that jets
in hadronic collisions are mostly produced by gluons, while in $e^+e^-$ they
come from the evolution of quarks. We should however point out that, as noticed
in a previous section, jets are defined according to different algorithms in
$e^+e^-$ and hadronic collisions, and unless a common definition is provided it
is not possible to draw quantitative conclusions from these comparisons.
Nevertheless it is encouraging that, as shown in Ref.~\cite{hubbard},
calculations based on the HERWIG Monte-Carlo are in agreement with the measured
inclusive fragmentation function.

\subsection{Multijet Final States}

	In all of the above, one has considered final states where two jets
predominate.  Predictions for multijet final states
are obtained either from QCD shower MC programs
or from fixed order PT.  For the
latter, one can obtain finite cross sections by limiting the minimum $E_t$
and opening angle of partons in order to stay away from soft
and collinear singularities.
For states selected where the partons are stiff and widely separated, one
expects that tree level predictions should be reasonably faithful.
There is no NLO calculation available for $N_{jet} > 2$, therefore
since one does not have the $\log(\mu)$ cancellation that appears at
NLO, there is a substantial sensitivity in the predicted
cross sections to a variation of the renormalization scale.
This is because the cross section is of order $\alpha_s(\mu)^N$ where $N$ is
the number of final state partons.  Any uncertainty in the scale will hence
be multiplied by a large factor in deriving cross sections.

The tree level matrix elements commonly in use  are based on calculations by
several groups\footnote{For a review of these techniques and for a complete set
of references, see Ref.~\cite{physrep}.}, and have been included in numerical
programs \cite{ks,bgk}\ which are currently used by the experiments.
Because of the complexity of the results, techniques have been developed to
provide reliable approximations  to these matrix elements
\cite{effpdf}-\cite{mpapprox}. The testing of these approximations using
current
data is very important, because rates for multi-jet production at the future
colliders will be extremely large and fast but reliable numerical simulations
will be required to evaluate them.

To start with, the topologies of multijet final states appear to be well
predicted by the tree level calculations.  Several examples can be seen in both
CDF and UA2 data.  CDF examined the topology of three jet events with high
statistics and in regions of uniform
acceptance. They found a very good agreement with tree level predictions
\cite{3jetprd}.  In particular, there is a distinct difference expected
between three jet topologies for events initiated by gluon-gluon
and gluon-quark collisions versus those from quark-antiquark collisions.
The data appear to be in good agreement with the expectation that most
of the three jet final states come from gluon-gluon and gluon-quark
collisions \cite{3jetprd}.

	Both UA2 and CDF explored in some detail the structure of
four jet final states \cite{ua4jets,cdf4jets}.
These studies are partly motivated by a search for double-parton
processes where two uncorrelated $2 \rightarrow 2$ scatters occur,
producing four jets in the final state.
Figure \ref{4jet} shows the angular separation of all pairings
of jets from four jet events in
CDF data \cite{cdf4jets} and compared with the results of
predictions based on the exact tree-level matrix elements \cite{ks},
which reproduce the data very well. In contrast to QCD production
where the four jets have no intrinsic correlation,
the double-parton events are expected to have jet pairs which approximately
balance in transverse momentum.
One expects that the cross section for double parton events
would scale like the square of the dijet ($2 \rightarrow 2$) cross
section, normalized by a factor that is comparable to the inelastic
cross section \cite{doublep}:
\begin{equation}
\sigma_{DP} = {\sigma_{dijet} \times \sigma_{dijet} \over 2 \sigma_{eff}}
\label{dps}
\end{equation}
where $\sigma_{eff}$ is expected to be roughly 10 mb \cite{cdf4jets}.
The factor of $2$ in the denominator is included to account for
the Poisson nature of chance of a double parton interaction
\cite{sjostrand}.
The Axial Field Spectrometer collaboration reported a significant
double-parton cross section, with $\sigma_{eff}=5$ mb \cite{afs4jets},  whereas
the UA2 collaboration did not find any evidence for the process and set a limit
of $\sigma_{eff} > 8.3$ mb (95~\% C.L.) \cite{ua4jets}\footnote{The
findings by UA2 and AFS are not necessarily inconsistent: on one side the
$x$-range probed by the multi-jet configurations  at the two energies of 63 and
630 GeV is very different. On the other, at the time of the AFS analysis the
exact predictions for the QCD four-jet production were not available.}.
Note that since $\sigma_{eff}$ appears in the denominator of
equation \eqnum{dps}, a larger number implies a smaller $\sigma_{DP}$.
Finally, CDF reported an effect at roughly a 2.5 standard deviation
level, with $\sigma_{eff} = 12.1^{+10.7}_{-5.4}$ mb.
It is estimated for four jet final states at the SSC, that
double parton scattering will dominate for jet $p_t$'s less than
40 GeV \cite{cdf4jets}.
If, in the future, a sizable effect is observed, it is
possible one may obtain unique information on correlations between partons in
the proton from double parton scattering.

The UA2 collaboration studied the cross section of  events with up to
six jets. Figure \ref{ua2multi} shows the jet $p_t$ distributions for 4, 5 and
6 jet final states, compared with various tree level predictions.
Notice that the normalization of the theory curves is absolute. Considering the
complexity involved in these calculations, the agreement with data is
remarkable and extremely encouraging in view of the potential applications of
these calculations to the study of multi-jet phenomena at the future hadronic
colliders LHC and SSC.

As an alternative to exact tree level calculations,  and to get a more
exclusive description of the events,  one may employ shower MC's.  In this
approach multiple jets can appear when branchings with large transverse
momentum relative to the leading partons take place. Given the approximations
involved in evaluating these large \pt\ branchings inside the MC,  an
assessment of the reliability of this approach is in order.

CDF recently performed a detailed comparison of the characteristics of
events with high total transverse energy with the HERWIG event generator
combined with a realistic detector simulation \cite{sumet}.
The events were selected by requiring that the total transverse
energy be in excess of 400 GeV.  Events with up to 6 jets were
observed, and the HERWIG generator does an impressive job in
reproducing a very large number of distributions, such as
the jet multiplicity as a function of different jet-$p_t$
thresholds (Figure \ref{jetmult}), the jet profiles,
invariant masses of various combinations of jets.
Such studies illustrate the power and accuracy of event generators to
reproduce event characteristics.

Although they agree in rough detail, there are some significant
differences among some of the MC event generators.  One
of the most relevant differences is how color flow is handled.
In particular, the dynamics of color flow leads to
the need of the angular ordering of QCD radiation in shower MC's
\cite{webberrev}.  The emission of radiation is related to
the color structure of the hard scattering process.  Some event
generators, such as ISAJET ignore the connection between radiation
and the hard scattering, whereas other generators, such as HERWIG
have explicitly built in the color flow connection to better
reproduce event properties.  The differences between the coherent
and incoherent emission has been studied extensively in $e^+e^-$ \cite{leprev},
but not as much in $p \overline p$ collisions.
This is largely
due to the inherent problems in distinguishing the soft particle flow
associated with the hard scattering from that associated with the
underlying event.  For sufficiently large momentum transfers, however,
one expects the radiation effects to become visible as jets, which
are more readily associated with the hard scattering process.
CDF studied the angular distribution of the third highest
$E_t$ jets in events with two high $E_t$ leading jets \cite{meschi}.
These studies show significant differences between the predictions
for ISAJET and HERWIG, where the data are in much better agreement
with HERWIG, indicating that for some measurements, color coherence
effects cannot be neglected.

\section{HEAVY FLAVOR PRODUCTION}
Heavy quark production in high energy hadronic collisions constitutes a
fundamental arena for the study of perturbative QCD.
Of particular importance is the role played by $m_Q$.   Only in
$b$ quark production does one have today the unique situation where
$m_Q \gg \Lambda_{QCD}$.   The prediction of heavy quark production
cross sections in hadronic collisions has far reaching implications.
Discovery reaches and limits for the ``top'' quark depend on reliable
estimates from PT.
The observability of CP violation in $B$ mesons \cite{cpviolation}\
at hadron colliders
depends, to a large extent, on the production cross section and
correlations between the $B$ and $\overline B$.
Recent years have witnessed remarkable progress both in the
theoretical understanding of the production mechanisms \cite{nason}\
and in the experimental capability to probe them via independent and
complementary observations \cite{nellis}.

The mass of the heavy quark $Q$ provides a natural infrared cutoff in the
evaluation of the production rates and multiplicities.  Complete NLO
calculations are available today for the total \cite{nde},
one-particle-inclusive \cite{nde_pt}\ and two-particle-inclusive \cite{mnr}\
cross sections.
Production of heavy quarks in the perturbative evolution of high energy jets
has also been studied, and LO expressions for the heavy quark multiplicities
are known \cite{binjets}.

The non-perturbative corrections which are required to derive the production
properties of observable heavy flavored hadrons $h_Q$ are suppressed by powers
of \lqcd/\mq. For production at large \pt, the factorization theorem guarantees
the existence of a fragmentation function $D_{h_Q}^{Q}(z,\mu)$ which  models
the fraction of momentum of the heavy quark retained by the heavy hadron:
\be \label{hvqfact}
	\frac{Ed^3\sigma_{h_Q}}{d^3p} \=
	\int \frac{E'd^3\hat\sigma_{Q}}{d^3p'} D_{h_Q}^{Q}(z,\mu)
	\frac{dz}{z^2},
\ee
where $p=zp'$ and $\hat\sigma_Q$ is the elementary cross section for the
production of the heavy quark Q, calculable as a perturbative expansion in \as.
The evolution of the fragmentation function with the factorization scale $\mu$
obeys the AP equation \cite{ap}\ with a boundary condition which
is given by $D_{h_Q}^{Q}(z,m_Q)=\delta(1-z)$,
up to non-perturbative effects \cite{hvqfrag}.
These non-perturbative effects obey a scaling law in \mq\ and can therefore be
parametrized in a phenomenological way by fitting, \eg, $e^+e^-$ data
\cite{peterson,colangelo}. With this additional input, non-perturbative
corrections to Equation~\ref{hvqfact}\ are suppressed by powers of \pt. The
evolution of $D_{h_Q}^{Q}(z,\mu)$ with $\mu$ is known today  up to NLO in
PT \cite{mele}.

When applied to the energy of the current hadron colliders,  these results are
believed to provide a reliable description of the production properties of very
massive quarks -- \eg\ the yet undetected $top$. In the case of {\it charm} and
{\it bottom}, the situation is more delicate. In fact
production of $c$ and $b$ quarks is dominated by gluon fusion processes ($gg\to
Q\bar Q$) and the distribution of gluons inside the proton is probed at values
of $x$ close to the boundary of current DIS measurements. Furthermore the
NLO contribution is larger than the LO result,  and very sensitive to the input
scale $\mu$.  Significant corrections are thus expected from yet higher order
terms. These corrections arise from a class of diagrams with $t$-channel
gluon exchange first appearing at NLO \cite{kunszt80} (Figure~\ref{fnlobot}).
They lead to
terms proportional to powers of $\as \log(s/\mq^2)$ \cite{nde}, which
might dominate at higher energies, as well as
becoming non negligible in the case of {\it top} production at
supercollider energies. Techniques exist to resum these large logarithms
\cite{lipatov}, and have been extended for application to this specific problem
\cite{smallx1,smallx2,smallx3}. Comparing the results of the NLO predictions
with the available data and verifying whether the resummed calculations can
explain possible differences is therefore of utmost importance as a test of QCD
per s\'e and as a milestone before extrapolation to higher
energies.

\subsection{Bottom Production}
There are several different channels which allow the detection of \b\ quarks.
Fully reconstructed exclusive decays of $b$-hadrons allow the unambiguous
tagging of a $b$-quark, together with a precise measurement of the hadron
momentum. Viable examples are provided by  $B^{\pm} \to \jpsi K^{\pm}$
\cite{ua1_Lb,cdf_psiK},  $B^{0} \to \jpsi K^*$ \cite{cdf_psiKstar}\ and
$\Lambda_b \to \jpsi \Lambda$ \cite{ua1_Lb}. Due to the small branching ratios
(BR) and detection efficiencies, these channels   are only accessible near
threshold ($\pt={\cal O}(\mb)$),  where the production rate of $b$ quarks is
more abundant.  The region of small \pt\ is expected to be more sensitive to
the uncertainties in the calculations mentioned previously and is therefore
potentially more interesting for critical tests of QCD.

At larger values of \pt\ (typically above 10$\div$15 GeV) semileptonic  decays
become the leading tool to study $b$ production.    Neglecting detector
backgrounds, and neglecting $W$, $Z$ and $c$ decays, $b$ quarks are the most
abundant source of high \pt\ leptons. Several techniques can be employed to
subtract the above backgrounds \cite{nellis}.  Backgrounds from $Z$'s, $W$'s
and continuum Drell-Yan events can be identified because single leptons from
these processes are more isolated than leptons from heavy quark decays,
surrounded by the fragments of a jet. In addition, lepton pairs from $Z$'s can
be eliminated with a cut on the invariant mass of the lepton pair, and $W$'s
can be identified by the large  transverse mass of the $\ell\nu$ pair.

For \pt\ values larger than 10$\div$15 GeV, the $c$ and $b$ cross sections are
comparable. Since $b$ quarks undergo a harder fragmentation into hadrons
compared to $c$ quarks, and since $B$ hadrons have a larger phase space
available for the decay, we expect  the $c$ contamination to contribute only a
fraction of the total lepton yield. This fraction can be precisely estimated by
studying the  transverse momentum of the lepton relative to the direction of
the  jet in which it is imbedded \cite{ua1_b}.

Furthermore, the $b$ component can be determined by tagging
charmed hadrons (say $D$'s) inside the jet and  with the correct charge
correlation with the lepton itself, \eg\ $e^-D^0$ as opposed to $e^-\bar D^0$.
UA1 has also pioneered a technique based on the detection of a
second lepton in the event \cite{ua1_b}. This second lepton comes either from
the charm emitted during  the decay of the $b$ into the leading lepton, or from
the decay of the second $b$ in the event. In the first case we have a low mass
dilepton pair, and the measured rate can be directly related to the $b$ cross
section. In the second case we have a high mass dilepton pair, and the
extraction of the inclusive $b$ cross section requires an understanding of the
correlations between the two heavy quarks in the event \cite{geiser}.

The advent of new technologies, such as secondary vertex detectors capable of
isolating the charged particles coming from the displaced vertex of a $B$
decay, will provide further tools to strengthen the capability of hadron
collider experiments to tag $b$ hadrons and study their properties.

Unlike other inclusive measurements (e.g. direct photon, jet production),
inclusive $b$ cross sections are reported as a function of the integral
cross section above some \pt.   This is done to minimize systematics
associated with the $b$ fragmentation and decay.  The effects
of these two effects must be unfolded in order to obtain a $b$ cross
section from the observed lepton spectrum.
The results of the measurements by UA1 \cite{ua1_b}\ and CDF
\cite{cdf_psiK}-\cite{cdf_bmu}\ are collected in Figure~\ref{fbpt}.
The two solid lines represent the NLO QCD prediction \cite{nde_pt}, obtained
using PDF's from the most recent MRS fit \cite{newmrs} (set D0) and two
different values for $\mu$ and \lqcd \footnote{It is worth pointing out that
the values of \lqcd\ extracted from fits to DIS data are systematically lower
than those obtained from precision measurements of jets performed at  LEP
\cite{lqcdLEP}. The differences are of the order of two standard deviations.
Using  for \as\ the values extracted by LEP experiments would increase the
predicted $b$ cross sections by an additional 20\%.}. This band is supposed to
represent an acceptable range of variation for the input parameters of the NLO
calculation.  The value of \mb\ was fixed to 4.75 GeV.  A variation of the mass
in the range 4.5 GeV $<\mb<$ 5 GeV only affects the result by no more than 20\%
in the region $\pt<10 GeV$, and of the order of few \% above 20 GeV. Two
features are to be noticed. First of all,  the theoretical uncertainty is
rather large, significantly larger than the uncertainties encountered in the
case of the NLO inclusive jet cross section. Secondly, while the  UA1 data fall
well inside the theoretical band, the CDF points are systematically higher,
with deviations of up to a factor of 3 for the low-\pt\ points.

No satisfactory explanation for this discrepancy is available as of
today, even though at least two suggestions have been put forward. First of
all the gluon momentum fractions $x$ probed by the CDF measurements are
significantly smaller than those probed by UA1. Attempts have been
made \cite{berger}\ to explicitly include the CDF $b$ data in global fits of
the gluon density. These
attempts have not led to a complete solution of the problem. An
explanation of this can be found in the following observation \cite{mlmb}: the
region in $x$ which is unexplored even by the most up-to-date DIS data is
$x<0.01$; using the available extrapolations of the gluon densities below this
value, the contribution to the cross section for $b$'s with $\pt>10$ GeV coming
from the region $x<0.01$ is only of the order of 20\%
(Figure~\ref{fcdflgx}).
Therefore only large differences in the extrapolation could explain the
observed discrepancy, and such differences are difficult to achieve because of
the global constraints posed by the measurements of gluon distributions at
larger values of $x$, such as momentum sum rules.

An alternative explanation could be provided by the presence of the large
$\log(s/m^2)$ corrections mentioned previously. The studies in
Ref's~\cite{smallx1}-\cite{smallx3}\ have led to a general reformulation of
the factorization theorem for application to processes where initial state
gluons with small momentum fraction $x$ are involved.  The result can be
expressed in terms of gluon distributions depending not just on $x$ and $\mu$,
but on the transverse momentum {\bf k} as well \cite{smallx1}:
\be \label{smallx}
	\sigma(s) = \int_0^1 dx_1 \int_0^1 dx_2
	\int_0^\infty d\mbox{\bf k}_1^2 \int_0^\infty d\mbox{\bf k}_2^2
	{\cal F}(x_1,\mbox{\bf k}_1,\mu){\cal F}(x_2,\mbox{\bf k}_2,\mu)
	I(\hat s,\mbox{\bf k}_1,\mbox{\bf k}_2),
\ee
where the functions ${\cal F}$ describe the transverse momentum distribution
of gluons with longitudinal momentum faction $x$. $I$, referred to in the
literature as the {\em impact factor}, represents the
gauge invariant
elementary cross section for the process $gg\to Q\bar Q$ with initial
off-shell gluons of virtuality $-\mbox{\bf k}^2$.
An intuitive physical interpretation of this result is the following: at small
$x$ and for $\mu\gg\lqcd$, gluons are more likely found in a peripheral branch
of the initial-state evolution tree. In other words, the multiplicity is
dominated by
processes where the degradation of the gluon momentum down to a fraction $x$
took place via a large number of successive splittings (see
Figure~\ref{fsmallx}).
Since at each splitting the gluon acquires some transverse momentum {\bf k},
{\bf k} will build up during the evolution to small-$x$; for $x$ small enough
the transverse momentum will not be negligible with respect to the scale of the
hard process, $\mu$. Therefore the description of the gluon density at
small-$x$ should depend on {\bf k} as well as on $x$ and $\mu$, and its
evolution  equation cannot neglect the transverse degree of freedom. An
evolution equation for the density ${\cal F}(x,\mbox{\bf k},\mu)$ can be
formulated \cite{lipatov}, extending the standard AP equation.
This evolution equation resums the leading $(\as\log(s/m^2))^n$ terms which
appear in the perturbative expansion for the hard scattering cross section and
allows them to be absorbed into ${\cal F}(x,\mbox{\bf k},\mu)$, provided one
uses the impact factor $I$ rather than the standard on-shell matrix element in
the expression for the cross sections, Equation~\ref{smallx}.
The result of this approach cannot be simply estimated by varying the
renormalization scale $\mu$  within some range, because the impact factor and
the
{\bf k}-dependent density contain information beyond what available in the
standard NLO calculation;
this could explain why even the change of $\mu$  in the rather extreme range
of $\mtr/4 < \mu < \mtr$ cannot reconcile the NLO prediction with the data.

The main physical consequence of this picture is that small-$x$  gluons
involved in a hard scattering at a scale $\mu$  will have an intrinsic
transverse momentum of the order of $\mu$  itself.  This additional transverse
momentum will smear the \pt\ distributions obtained from a pure NLO
calculation, but complete calculations of this effect are not yet available.
Explicit estimates exist \cite{smallx1}\ of the  corrections to
the total cross section resulting from Equation~\ref{smallx}. At Tevatron
energies these corrections amount to approximately 50\% of the NLO total cross
section. While this effect seems insufficient to explain the observed
discrepancy, one should keep in mind that the smearing induced by the effective
intrinsic \pt\ introduced by Equation~\ref{smallx}\ could very well push most
of this contribution to values of $\pt>\mb$, where the NLO cross section is
only a fraction of the total.

While we await for more calculations, it is worth exploring  additional
consequences of this scenario. In addition to  pushing the measurement of $b$'s
to smaller values of \pt, it is useful to study correlations between the $b$
pair. NLO calculations exist for these correlations \cite{mnr}. If the
small-$x$ effects behave as indicated, we would expect to observe a flattening
of the $\Delta\phi$ and $\pt^{b\bar b}$ distributions relative to the NLO
prediction. Here $\Delta\phi$ represents the difference in azimuth between the
$b$ and the $\bar b$, and $\pt^{b\bar b}$ represents the transverse momentum of
the pair.  The flattening would be caused by the additional intrinsic \pt\ due
to the gluon transverse momentum {\bf k}.

The $\Delta\phi$ correlations have been studied by UA1
\cite{geiser}, indicating a good agreement with the NLO calculation.
This result does not resolve the issue, however, because the agreement of the
NLO $b$ cross section with the data suggests that the energy at UA1 is below
the threshold for the possible onset of these new small-$x$ phenomena.

\subsection{Charmonium Production}
In this Section we review the status of the measurements of production
cross sections for charmonium resonances such as the \jpsi.
The theory of quarkonium production \cite{onia}\ is on a less solid ground than
the  theory of open heavy-quark production. Production cross sections are
evaluated
by convoluting the $c\bar c$ matrix elements with the non-relativistic
charmonium wave function, parametrized in terms of the decay widths of the
relevant $(J,L)$ state. The QCD radiative corrections to the LO processes have
not been evaluated yet.

The observation of \jpsi's is however an important ingredient in the study of
$b$ production. On one hand, a significant fraction of the detected
\jpsi's  comes directly from $b$-hadron decays rather than from prompt
charmonium formation \cite{gms,ua1_psi}. In fact the \jpsi\ form factor
inhibits production  with $\pt\gg m_c$.
On the other hand, $b$-decay final states involving a \jpsi\ provide
unique tags in the search of yet unobserved or rare $b$-hadrons (such as $B_s$,
$B_c$, $\Lambda_b$) as well as in the detection of CP asymmetries (\eg\ from
$B_d \to \jpsi K^0_{\rm S}$ decays \cite{cpviolation}.)
A coherent picture of the production of both $b$ and \jpsi\ in
hadronic collisions will therefore provide not only a significant test of QCD,
but also the starting point for important studies of the Standard
Model.

Figure~\ref{fpsipt}\ shows the inclusive \pt\ differential distribution for
\jpsi's measured by UA1 \cite{ua1_psi} and CDF \cite{cdf_psi}.
We superimpose the result of a QCD calculation \cite{mlmb}\ based on the LO
matrix elements given in Ref.~\cite{onia}\ for the direct charmonium
production,
plus the contribution from $B$ decays evaluated using NLO matrix elements
\cite{nde_pt}, convoluted with a Peterson fragmentation function and the
experimentally observed $B\to\jpsi$ decay spectrum.
The theoretical error band is evaluated using the same range of parameters
\lqcd\ and $\mu$ employed before in the study of the $b$ cross sections.
Notice that changing $\mu$  for the direct charmonium contribution
causes a variation ranging from a factor of 7 to 10, depending on \pt.
This indicates that the LO prediction for direct charmonium is very poor, and
very large NLO corrections should be expected.

In the case of UA1 the data fall all inside the theoretical band, while again
CDF shows a production rate larger than expected. A similar feature is observed
in the CDF measurement of the \psitwos\ \pt\ distribution \cite{cdf_psi}.

An important parameter is the fraction of \jpsi's coming from $b$ decays,
$f_B$. This number allows to extract a $b$ cross section from the observed
\jpsi\ production rate. Notice from the theoretical curves in
Figure~\ref{fpsipt}\ that $f_B$ is very sensitive to the parameters used for
the evaluation of the two contributions.

$f_B$ can be extracted
experimentally, for example by separating the direct \jpsi's from those due to
$B$ decays via the observation of the displaced vertex from which the \jpsi\
originates, due to the long $B$ lifetime. UA1 measured $f_B$
(32\% for $\pt(\psi)>5$ GeV \cite{ua1_psi}) by assuming that
direct \jpsi's are isolated while \jpsi's from $B$ decays are not.
This number is consistent with the estimates provided in \cite{mlmb}.

The
assumption used by UA1 to extract $f_B$ might not be correct if other
production mechanisms were responsible for direct quarkonium production, such
as for example gluon $\to$ \jpsi\ fragmentation \cite{braaten}. It is
reasonable to expect that at some value of \pt\ the dominant production
mechanism for charmonium states will indeed be via gluon fragmentation. The
main reason being that direct production as described by the LO mechanisms
inhibits production at large \pt\ via a form factor suppression (the
probability that a charmonium bound state will hold together when produced
{\em directly} in an interaction with a large virtuality scale is highly
suppressed). The fragmentation functions for the creation of $S$-wave
charmonium ($\eta_c$ and \jpsi) in a gluon shower have recently been calculated
\cite{braaten}\ and those for the creation of $P$-wave states ($\chi$) will
soon be available (E~Braaten \&\ TC~Yuan, personal communication).

These calculations can be used to extract the fragmentation contribution to
charmonium production in the regions of \pt\ explored experimentally, and to
verify whether this process can account for the large
observed rates. The experimental detection of non-isolated \jpsi's from a
primary vertex, therefore not coming from $B$ decays, would
indicate that these processes are indeed present.

Measurements of the decay-vertex position of the \psitwos\ would
provide evidence in favour or against the current belief that most of them come
from $B$ decays. If the gluon fragmentation mechanism were important, it
would appear with a signal of non-isolated prompt \psitwos.

Similarly interesting would be a measurement of the $\chi$ \pt\
spectrum, which is expected to be dominated by direct production rather than
$B$ decays. A preliminary measurement by CDF \cite{cdf_chi} reports
$BR(\psi\to \mu^+\mu^-) \times$  $\sigma(\chi_c\to\psi\gamma;\; \pt_\chi > 7
GeV; \; \vert \eta \vert<0.5)$ = $3.2\pm0.3\pm1.2$ nb. Both $\chi_1$ and
$\chi_2$ are here included.  This can be compared with the range  $0.64 nb <
\sigma < 5.1 nb$ obtained using the LO QCD calculation described above
\cite{mlmb}. Using the above cross section and using the inclusive $B\to
\chi_{c1}$ branching ratio of $0.54\pm0.21$\% \cite{bchibr}, we estimate that a
fraction smaller than 10\% of the $\chi$'s comes from $B$ decays.

A measurement of the production cross section and \pt\ spectrum for $\Upsilon$
states would also be very useful in understanding the quarkonium production
mechanisms. In this case one has at least three advantages: (i) the masses
involved are larger and presumably both the non-relativistic approximation
involved in the determination of the quarkonium wave function and QCD PT would
work much more reliably than for charmonium; (ii) the signal does not have a
contamination similar to the one due to $B$ decays; (iii) the \pt\ spectrum
could be extended to very small values of \pt, thanks to the large mass of the
$\Upsilon$ and the large momentum of the decay muons.
\section{$W$ AND $Z$ PRODUCTION}
\subsection{Inclusive Measurements}
Inclusive production of \w\ and \z\ bosons is the most accurately known process
in hadronic collisions. The absence of final state strong interactions
affecting the observed state, one or two large-\pt\ charged leptons,
allows for high precision measurements and calculations.
Uncertainties
in the measurement of the total cross sections
\cite{sps_wxsec,cdf_wxsec}\ are less than
10\% and are dominated by the uncertainty on the absolute luminosity
(see Table~\ref{twzxsec}).
{\renewcommand{\arraystretch}{1.2}
\begin{table}
\begin{center}
\caption[]{  \label{twzxsec}
\rightskip=2pc\leftskip=2pc\baselineskip=12pt
$\sigma_W \cdot BR$, $\sigma_Z \cdot BR$ and $R=\sigma_W \cdot BR/\sigma_Z
\cdot BR$ at 630 and 1800 GeV.
Data vs. \orderas{2}\ QCD for different PDf sets \cite{wxsect}.
BR($W\to \ell\nu_\ell$)=0.109 and BR($Z\to \ell^+\ell^-$)=
3.35$\times 10^{-2}$. }
\begin{tabular}{llccc} \hline
 & Data & HMRSB & MTE & MTB \\ \hline\hline
$\sigma_W \cdot BR$ (pb) 630 GeV & UA1: $609\pm 41\pm 94$ & & & \\
&             UA2: $682\pm12\pm40$ & 733 & 699 & 720 \\ \hline
$\sigma_Z \cdot BR$ (pb) 630 GeV & UA1: $58.6\pm 7.8\pm 8.4$ & & & \\
&             UA2: $65.6\pm 4.0\pm 3.8$ & 69.2& 71.0& 69.9\\ \hline
$R$ (630 GeV) & UA1: $10.4^{+1.8}_{-1.5}\pm 0.8$ & & & \\
&               UA2: $10.4^{+0.7}_{-0.6}\pm 0.3$ & 10.6 & 9.9 & 10.3 \\ \hline
$\sigma_W \cdot BR$ (nb) 1800 GeV &
              CDF: $2.20\pm 0.04\pm 0.20$ & 2.06 & 2.02 & 2.10 \\ \hline
$\sigma_Z \cdot BR$ (pb) 1800 GeV &
              CDF: $214 \pm 11  \pm 20  $ & 194  & 192  & 198  \\ \hline
$R$ (1800 GeV) &
              CDF: $10.0\pm 0.6 \pm 0.4$ & 10.6 & 10.5& 10.6 \\ \hline\hline
\end{tabular}
\end{center}
\end{table} }
The full NNLO \orderas{2}\ corrections to the cross section are known
\cite{wxsect} and techniques for the resummation of classes of leading and
subleading logarithmic corrections to all orders of PT are available
\cite{dyresum}. The current theoretical systematic error is below 5\%,
estimated by varying factorization and renormalization scales within the range
10 GeV $<\mu<$ 1000 GeV. Slightly larger uncertainties arise from the use of
different PDFs. The agreement between theory and experiment, at both S$p\bar
pS$ and Tevatron energies, is within one standard deviation and does not favour
any particular set of PDF's provided one uses recent NLO fits.
Even though the \orderas{2}\ corrections add only a very small numerical
contribution to the \oas\ result, they conspire to improve the
stability of the cross section under changes of $\mu$ by a factor of 3-5,
depending on the beam energy and PDF set \cite{wxsect}.  This stability and the
agreement with data represent a remarkable success of perturbative QCD.

The charged lepton rapidity
asymmetry in \w\ decays:
\be
	A(y)=
    \frac{d\sigma/dy(\ell^+)-d\sigma/dy(\ell^-)}
         {d\sigma/dy(\ell^+)+d\sigma/dy(\ell^-)}.
\ee
is more sensitive to the choice of PDF set and is not affected by luminosity
uncertainties. Its measurement probes directly the quark components and the sea
flavour symmetry of the proton \cite{wasy}, necessary ingredients for a precise
measurement of the $W$ mass \cite{kevin}.
Current data at  $\vert y \vert < 2$  \cite{cdf_wasy}\ already discriminate
between different PDF fits. The \oas\ calculation of this asymmetry is
available \cite{reno_wasy}, and new data will hopefully extend
this measurement to more forward regions, where the difference between PDF's is
expected to be even more pronounced.

NLO calculations have also been recently completed for the inclusive \w\ and
\z\ \pt\ distributions \cite{reno_wpt}.  Measurements have been reported by
UA1, UA2 and CDF \cite{sps_wpt,cdf_wpt}, and are shown in Figure~\ref{fwpt}.
The main source of systematic uncertainties in the case of the $p_t^W$
measurement is the determination of the neutrino transverse momentum, degraded
by the energy resolution for the jets possibly present  in the event. The small
statistics (10\% relative to the $W$ case) limits instead  the otherwise very
clean $p_t^Z$ measurement.

The agreement with QCD is good at large \pt, indicating consistency with the SM
expectations. At smaller \pt\ the theory is in better agreement with the UA2
data than with CDF. The small \pt\ region is interesting from the theoretical
point of view, because a correct description of the spectrum requires the
resummation of multiple gluon emission, which can be calculated in perturbative
QCD \cite{sudakov}\ in the form of Sudakov form factors \cite{PQCD}.
These effects have been included in the theoretical curves shown here
\cite{arnold_wpt}\ using the techniques developed in \cite{wpt_resum}.
Additional higher statistics measurements of the $Z$ \pt\ spectrum will help
turning the qualitative agreement indicated here into solid QCD tests in the
delicate semi-inclusive $\pt\to 0$ region.

\subsection{Associated Jet Production}
The production of jets associated with $W$'s and $Z$'s is
less well predicted than the inclusive momentum spectra.  Nonetheless,
the characteristics of multijet final states in these events is very
topical since it forms a background to top production. As with
purely hadronic final states, most predictions for multijet
characteristics in $W$ and $Z$ events are only available at tree
level \cite{treew}, hence absolute cross section estimates have large
uncertainties associated with the $\alpha_s^N$ terms.  Recent work \cite{giele}
has led to new NLO
predictions for quantities such as the jet $E_t$ and pseudorapidity
distributions in W+1 jet events.

CDF and UA1 have measured the multiplicities of jets associated with $W$ and/or
$Z$ production and have compared the results to tree level predictions
\cite{cdfwjets,ua1wjets}.   Within the relatively large statistical and
theoretical uncertainties the results are in good agreement with the theory.
Figure \ref{wnjet} shows the cross section for $W$ production as a function of
jet multiplicity from CDF data.   Other distributions, such as the $E_t$
distribution of associated jets, show some discrepancy with tree level
predictions \cite{dallashuth}.  New NLO predictions will possibly improve the
agreement with the data \cite{giele}. With more data at the Tevatron, it is
expected that a more thorough test of $W$ and $Z$ plus jet production can be
carried out.

Using the ratio of the \w+1 jet and \w+0 jet event rates, and comparing with
the results of a LO calculation for \w+1 jet production, UA2 has extracted
a measurement of \as($M_W^2$):  \as=$0.123\pm0.018(\mbox{
stat.})\pm0.017(\mbox{syst.})$.  This value is consistent with other
determinations of \as\ from LEP and DIS data \cite{lqcdLEP}. We point out that
a fully consistent measurement of \as\ and an extraction of \lqcd\ can however
only be performed using a NLO calculation for the \w+1 jet process. Only at
this order it is possible to reduce the $\mu$ scale uncertainties and to define
a precise renormalization scheme within which \as\ is measured.
New  analyses using the calculations of Ref.~\cite{giele}\ will hopefully
follow.
\section{DIRECT PHOTONS}
\hyphenation{brems-strahl-ung}
\subsection{Single Photon Production}
As in the case of Drell Yan, the measurement of photons produced directly in a
hadronic collision \cite{gammarev}\ has the advantage of not suffering from
final state strong  interactions. Furthermore, since EM energy is detected with
much better resolution than hadronic energy, systematic errors in the
measurement of the photon momentum and direction are smaller than in jet
measurements.
Production of direct photons at small \pt\ is dominated by processes with a
$qg$ pair in the initial state, be them of the Compton
or of the bremsstrahlung type (Figure~\ref{fphodia}).
The capability of the
experiments to observe direct photons at small \pt\ provides therefore yet
another potential tool, in addition to the $b$ quark measurements,
to explore the gluon content of the proton at small
values of $x$, or alternatively to learn more about small-$x$ phenomena.
The associate production of photons and charm quarks has
also been suggested as a direct probe of the charm density in the
proton \cite{chapdf}.

Several difficulties however complicate the study of direct photons.
First of all there are severe backgrounds to photon identification coming from
hadrons such as $\pi^0$ and $\eta$'s decaying into almost collinear photon
pairs, faking a single $\gamma$.  This background is statistically subtracted
using two techniques. One technique relies on the different probability that
one
photon or a photon pair will convert in a $e^+e^-$ pair, the probability being
independent of \pt. This ``conversion method'' can be used for arbitrarily
large values of \pt. The second technique relies on the measurement of  the
transverse shape of the EM shower in the calorimeter  to determine the fraction
of events with two overlapping photons. This ``profile method'' can only be
applied over a  limited \pt\ range, above which the two photons are too close
to be separable.

On the theoretical side, predictions depend on the
knowledge of bremsstrahlung contribution, which has both a perturbative and a
non-perturbative piece. The latter is needed to properly define the boundary
condition of the perturbative parton$\to$ photon fragmentation function. It is
due to the intrinsic hadronic component of the photon and it leads
to a non-negligible $g\to$ $\gamma$ fragmentation  probability via
Vector Meson Dominance (VMD).

To reduce the hadron decay backgrounds, experiments do not measure a fully
inclusive spectrum, but the so called isolated photon spectrum. Isolation is
defined in different possible ways. UA2 requires no charged tracks within a
$\Delta\eta \times \Delta\phi=0.2\times 15^\circ$ window around the $\gamma$
direction, and no EM energy within $\Delta R<0.265$. CDF requires the presence
of less than 2 GeV of hadronic energy inside a cone of radius $\Delta R<0.7$
surrounding the photon.
The isolation reduces the bremsstrahlung
contribution \cite{iso_pho}-\cite{pho_zolly}\ and emphasizes the
purely perturbative effects, allowing for a more direct test of QCD.

Full NLO calculations are available for the inclusive
\cite{inc_pho}\ and isolated \pt\ spectrum
\cite{iso_pho,pho_nlofrag}, as well as for the photon+jet processes
\cite{jet_pho}.  A
detailed study of the effects of isolation is presented in \cite{berger_iso}.
The comparison between theory and data is shown in Figure~\ref{fonepho}, which
includes both UA2 and CDF results.
While the agreement for $\pt>20$ GeV is
rather good, a discrepancy is apparent at smaller \pt\ values. This is even
more clear at the Fermilab energy. Several effects could be responsible
for this problem. We will briefly survey them here.

First of all, as always in PT, there is an intrinsic
scale uncertainty. Here the scales needed are three: for renormalization,
initial state factorization and final state fragmentation. Studies reported in
\cite{cdf_pho}\ indicate that the shape of the spectrum is rather insensitive
to the scale uncertainty, at least in the \pt\ range probed experimentally. Not
even the use of different PDF sets can accommodate the factor of 2 discrepancy
observed for the lowest \pt\ bins \cite{cdf_pho}. As in the case of the \b\
cross section, Figure\ref{fcdflgx}, the values of \pt\ are
probably too large to allow significant departures from current PDF fits.

The next possible effect is the bremsstrahlung contribution: how well do we
know it? Ref.~\cite{pho_nlofrag}\ describes the full NLO correction to the
bremsstrahlung processes, including a VMD description of the photon as a
phenomenological input for the evaluation of the $g\to\gamma$ fragmentation.
The results indicate that higher order terms add at most 50\% to the
lowest order fragmentation contribution to the inclusive spectrum. After
isolation cuts their effect will be even smaller, because the $g\to$
isolated-$\gamma$ fragmentation is highly suppressed. We believe that 50\% is
therefore a reasonable estimate of the uncertainty reached today on the size
of the bremsstrahlung contribution.
Figure~\ref{fphoiso}\ shows the effect of removing the isolation requirement
from the NLO QCD calculation \cite{cdf_pho}. This increases the QCD
result by no more than 30\%. A 50\% uncertainty on this number is not
sufficient to entirely explain the observed differences.

We cannot exclude that a combination of all three effects just
considered, in addition perhaps to new data and a better understanding of
the experimental systematics, can reestablish agreement between theory and
observations. Another possibility is however open.
That is, the violation of naive factorization at small $x$, as was discussed in
the heavy quark section. Like in that case, new diagrams with a
$t$-channel gluon exchange appear at NLO for the first time
(Figure~\ref{fphodia}). The same
considerations and techniques outlined previously apply to this case
\cite{smallx1}, even though no explicit calculation of the corrections to the
differential \pt\ spectrum has been carried out as yet. This issue will have to
be properly understood before the photon distributions -- either in \pt\ or in
rapidity -- can be used to extract sensible measurements of the gluon structure
functions in the small $x$ region \cite{iso_pho,ua2_pdf}.

\subsection{Double Photon Production}
Interesting measurements have also been performed on the direct production of
photon pairs. Aside from its interest for QCD, this process is undergoing
intense scrutiny as a possible dominant source of background to the detection
of an intermediate mass Higgs boson at supercollider energies \cite{lhc_dipho}.
The capability of QCD to properly estimate the $\gamma\gamma$ production rate
is therefore a very important fact to establish.

Three processes contribute to the production of $\gamma$ pairs
(Figure~\ref{fdiphod}): direct quark
annihilation ($q\bar q\to \gamma \gamma$, ${\cal O}(\alpha^2)$), gluon fusion
via a quark box diagram ($gg\to \gamma \gamma$, ${\cal O}(\alpha^2\as^2)$) and
various bremsstrahlung contributions ($qg\to q\gamma \gamma$,
${\cal O}(\alpha^2\as)$).
Even though of different order in \as, these
contributions are all comparable in magnitude over the currently measured \pt\
range, because at small $x$ we have $q(x) \sim \as g(x)$. The complete  ${\cal
O}(\alpha^2\as)$ calculation is available \cite{nlo_digamma},  including the
effect of isolation cuts \cite{owens_digamma},  together with the LO $gg\to
\gamma\gamma$ process. Data from UA2 \cite{sps_pho}\ and CDF \cite{cdf_dipho}\
are shown in Figure~\ref{ftwopho},
compared to the relative calculations.
In the case of UA2 the photons are not
required to be isolated. Backgrounds and bremsstrahlung are reduced by applying
the cut: $\vec{\mbox{\bf p}}_t(\gamma_1) \cdot \vec{\mbox{\bf p}}_t(\gamma_2) <
-0.7 \vert \vec{\mbox{\bf p}}_t(\gamma_1) \vert^2$. The theory calculations
reproduce the experimental selection criteria.

The CDF data are systematically above the QCD curve, in particular at low \pt.
UA2 shows a discrepancy only in the first \pt\ bin. In addition to the pure QCD
curve, the figure shows the results obtained by the PYTHIA shower MC, with and
without the bremsstrahlung terms. The comparison between the different curves
suggests that $i)$ PYTHIA has a bremsstrahlung contribution larger than NLO QCD
and $ii)$ initial state radiation induces a significant smearing of the \pt\
spectrum. It is perhaps premature to formulate a judgement in relation to this
measurement. On one side the statistical errors are still large. On the other
the calculations have not been completed at the full  ${\cal
O}(\alpha^2\as^2)$, where we know some important contributions ($gg\to
\gamma\gamma$) but we ignore the effect of others a priori comparable in size,
such as $gg\to q\bar q \gamma\gamma$. This last process would also contribute
to a broadening of the $\gamma\gamma$ correlations w.r.t the available ${\cal
O}(\alpha^2\as)$ estimates, which are unable to explain the data
\cite{cdf_dipho}.
Last but not least, the values of $x$ probed by this measurement are even
smaller than those relevant for the $b$ cross section, therefore this process
is another interesting candidate for the study of small-$x$ effects on
production mechanisms.

CDF also measures the average transverse momentum of the photon pair, $\langle
K_t \rangle =5.1\pm 1.1$ GeV. This is consistent with what expected from
perturbative initial state radiation, $\langle
K_t \rangle \sim \as \langle \sqrt{\hat s} \rangle \sim 4$ GeV, considering
that the bremsstrahlung processes will contribute an additional unbalance.
CDF quotes agreement with the prediction of the PYTHIA calculation for
$\langle K_t \rangle $.

\section{CONCLUSIONS}
All hard scattering processes in hadronic collisions require some
understanding of QCD to be properly described.  This is valid
to the extent that
they depend explicitly on $\alpha_s$ and the parton distribution
functions.  Although QCD is widely accepted as the theory of strong
interactions, progress can only result from successively making
more rigorous
tests, where discrepancies are not idley dismissed,  but both data and
theoretical assumptions are closely examined.

	In order to summarize the status of QCD predictions, one can
imagine two ways of classifying results.  In the first one could
select phenomena according to the quality of the agreement between
theory and experiment.  In the second, one can select phenomena
according to the presumed reliability of theoretical predictions and
the corresponding faith in experimental results.  It is a fact that
processes which are believed to be reliably calculated also happen to
belong to the class for which the agreement with data is good.
This is the case for the 1-jet inclusive distributions and for
$W$ and $Z$ production, which should be considered as successes of
the application of perturbative QCD to hadronic collisions.
There is, however, a possible discrepancy in $x_T$ scaling for
jets, which should be an incisive test for the theory.  As this
article goes to print, there is no obvious explanation for such
a discrepancy and we look forward to resolution, either via more
data, or a new insight in the comparison to theory.

	In contrast to inclusive jet and  $W$,$Z$ production, there are
processes such as $b$ quark and direct photon production, where the
theoretical uncertainties are large even at NLO.  Perturbative $K$ factors are
big and strongly dependent  on the choice of factorization and renormalization
scales.  Even  worse, the disagreement between theory and data seems to be
larger than the presumed uncertainties can account for.   With independent data
for parton distributions in this range of $x$, it appears unlikely that one can
find  fault in a lack of knowledge of the  gluon densities.   There are, on the
other hand, strong indications that a deeper understanding of the perturbative
picture may be required to explain the discrepancies. In the case of $b$ cross
sections, more data, particularly with the power of secondary
vertex detectors, will provide strong checks on the data.

As indicated in the review of direct photon results, the
processes contributing to photon or heavy quark production at the
NLO have singularities which are not present at tree level.  For
example, this is the case of diagrams with a $t$-channel exchange.  Since
these singularities only appear at NLO, an even higher order calculation
would be needed to have a true NLO approximation to all relevant processes.
This does not represent a problem for the 1-jet inclusive distributions
or for the $W$, $Z$ and Drell-Yan: in the first case no new
singularity appears at NLO ($t$-channel gluon exchange is already there at
tree level), in the second case, the available calculations are
already at NNLO.   This distinction could explain why there appears
to be two classes of processes.

Perturbative techniques for the study of multijet configurations are
rapidly evolving and the agreement with data is quite reasonable.
These tests are crucial to the search for new phenomena in events
containing multiple jets.

The measurement of finer details of the event structure, such as jet shapes,
fragmentation and multijet correlations  shows a good agreement with the
results
of both shower MC's and  parton level calculations. This is therefore a success
of perturbative QCD and of the way higher order processes are included in the
MC algorithms.  These measurements support the concept of local
parton-hadron duality and establish a firmer ground for the use of shower MC's
to predict the fine details of the jet structure in hadronic collisions.

\vskip 1cm
{\bf Acknowledgments} We wish to thank the members of UA1, UA2 and CDF for
making their results available for this review, in particular D. Wood, P.
Lubrano and KH Meier. In the preparation of this review we benefited from
discussions  with S. Behrends, S. Ellis, RK Ellis and Z. Kunszt.


\def\figlist{\section*{FIGURE CAPTIONS\markboth
 {FIGURECAPTIONS}{FIGURECAPTIONS}}\list
 {Fig. \arabic{enumi}:\hfill}{\settowidth\labelwidth{Fig. 99:}
 \leftmargin\labelwidth
 \advance\leftmargin\labelsep\usecounter{enumi}}}
\let\endfiglist\endlist \relax

\begin{figlist}
\item{\label{fnewpdf}
Gluon densities according to the most recent PDF analyses:
MRSD0 \cite{newmrs}\ and various CTEQ fits \cite{CTEQ}. For $0.01<x<0.1$ and
$Q>10$ GeV, differences never exceed the 10\% level.}

\item{
Comparison of the inclusive jet cross section for
LO QCD predictions with experiments at the ISR,
S$p \overline p$S and Tevatron colliders \cite{cdfjets,ua2jets,isrjets}.
Only one free parameter
in the theory has been fixed (renormalization scale) in order to
obtain this figure.\label{locompar}}
\item{
Examples of diagrams contributing to
the jet cross section at ${\cal O}(\alpha_s^3)$.
Collinear and soft singularities cancel between
loop and tree diagrams, after imposition of a sensible jet definition
involving finite opening angles for the final state partons.\label{nlodiag}}
\item{
Sensitivity of the inclusive jet cross section to
the renormalization scale for ${\cal O}(\alpha_s^2)$ and
${\cal O} (\alpha_s^3)$ predictions.  Note that near the
hard scattering scale, $E_t$, the sensitivity is greatly reduced for
the NLO calculation and the $\mu$ dependence goes from
monotonic to forming a plateau near the hard scattering
scale. \label{renorm}}
\item{
The inclusive jet $E_t$ spectrum for CDF data using
a cone size of $0.7$, compared to theory as a ratio of
(Data -- Theory)/Theory.   The upper plot (a) illustrates the
theoretical uncertainty associated with variation of
the renormalization scale $\mu$ ($E_t > \mu > E_t/4)$ for
both LO and NLO. The lower plot (b) illustrates
the dependence on the choice of PDF. The
${\cal O}(\alpha_s^3)$ prediction using the HMRS set B \cite{hmrs}
PDF is used as a reference.\label{linear}}
\item{
The variation of the jet cross section with clustering cone size
$R$ for jets of 100 GeV $E_t$.
The standard ${\cal O}(\alpha_s^3)$ calculation uses the
merging parameter $R_{sep}=2.0$, whereas a modified version employs
$R_{sep}=1.3$ \cite{eksprl}.
\label{conevar}}
\item{
Fraction of energy contained in a sub-cone of radius $r$
in jets found with a R=1 cone. The data are from
CDF charged tracking information, the QCD predictions are from
\cite{eksprl} and HERWIG \cite{herwig}.
\label{flow}}
\item{
The ratio of dimensionless cross sections measured
at $\sqrt{s}=$ 1.8 TeV and 0.546 TeV compared to QCD predictions
at both LO and NLO. \label{xt}}
\item{
The dijet angular distribution, $dN / d \chi$ from
CDF data \cite{mueller} shown
along with ${\cal O}(\alpha_s^2)$ and ${\cal O}(\alpha_s^3)$
predictions.  The data are divided into three bins of dijet
invariant mass, $M_{jj}$.\label{angular}}
\item{
The dijet angular distribution from CDF data \cite{mueller}
for the highest values of $M_{jj}$ compared with a model which includes
both QCD and a parameterization for the effects of quark compositeness.
\label{compositeness}}
\item{
The evolution of the jet fragmentation function, $F(z)$,
as a function of dijet invariant mass, $M_{jj}$.  This is shown
along with fits of the form $A \, ln M_{jj} + B$ \cite{hubbard}.
\label{evol}}
\item{
Angular separation for pairs of jets in four jet events.
The solid line are the predictions from exact LO QCD
matrix elements \cite{ks}, and the
dashed line represents the expectations of phase space.  The tree
level predictions clearly describe the data much better than phase
space. Jets are ordered by $p_t$\cite{cdf4jets}. \label{4jet}}
\item{
The distribution of jet $p_t$  for 4, 5 and 6 jet
events from the UA2 collaboration \cite{ua4jets}.
The solid curve represents the exact LO QCD calculation \cite{ks,bgk} for four
jets. The dashed-dotted line is the result of the Maxwell approximation
\cite{maxwell} for five jets, and the dashed lines are the predictions using
the Kunszt-Stirling approximation \cite{ksapp}.  \label{ua2multi}}
\item{
\label{jetmult} The jet multiplicity plotted for
different minimum jet $p_t$ cuts for events with greater than
400 GeV total transverse energy from CDF data \cite{sumet}.
The histograms are from the
HERWIG event generator combined with a detector simulation.  Each
histogram represents a different choice of PDF;
employed were DO1 (solid), DO2 (short-dashed) \cite{dukeowens},
EHLQ1 (long dashed) and EHLQ2 (dot-dashed) \cite{ehlq}.}
\item{\label{fnlobot}
A representative diagram for the $t$-channel gluon exchange contribution to
heavy quark production.}
\item{\label{fbpt}
Integrated $b$ \pt\ distribution at UA1 (left) and CDF
(right): data versus NLO QCD.
The lower curves correspond to $(\mu,\lqcd)=(\mtr,215$ MeV), the upper ones
to $(\mu,\lqcd)=(\mtr/4,275$ MeV), with $\mtr^2=\pt^2+\mq^2$. 275 MeV
corresponds to one standard deviation from the central value of the MRSD0 fit
for \lff.}
\item{\label{fcdflgx}
Fraction of the NLO QCD $b$ cross section at 1.8 TeV coming from gluons with
$x_g<x$, for different \pt\ thresholds.}
\item{\label{fsmallx}
A picture of the evolution of a gluon towards small-$x$.}
\item{\label{fpsipt}
$J/\psi$ \pt\ distribution at UA1 (left) and CDF  (right): data versus
QCD. Dotted line: direct quarkonium, dashed line: $b$ decays, solid: total. The
lower set of curves correspond to $(\mu,\lqcd)=(\mtr,215MeV)$, the upper set to
$(\mu,\lqcd)=(\mtr/4,275MeV)$. Parton distribution set MRSD0 \cite{newmrs}.}
\item{\label{fwpt}
$W$ \pt\ distribution at UA2 (right) and CDF (left): data versus
QCD \cite{arnold_wpt}. The band indicates the theoretical uncertainty due to
the choice of factorization scale and PDF sets.}
\item{\label{wnjet}
\w+ $n$ jet production rates at $\sqrt s =1.8$ TeV \cite{cdfwjets}.}
\item{\label{fphodia}
Sample diagrams contributing to prompt photon production. Left: LO Compton
scattering; Right: NLO bremsstrahlung.}
\item{\label{fonepho}
Isolated prompt photon \pt\ distribution at CDF and UA2, compared to
a NLO QCD calculation \cite{iso_pho}. For CDF, profile (circles) and
conversion (diamonds) methods have separate normalization uncertainties,
shown in the legend. }
\item{\label{fphoiso}
Study \cite{cdf_pho}\ of the effect of isolation on the photon \pt\ spectrum at
1800 GeV. The solid lines indicate the relative variation of the theoretical
calculation after reducing the isolation cone to 0.4, and after removing the
isolation.}
\item{\label{fdiphod}
 Sample diagrams contributing to double prompt photon production.}
\item{\label{ftwopho}
 Double prompt photon \pt\ distribution at 630 and 1800 GeV,
compared to various theoretical calculations. The \pt\ of both photons in each
event enter in the plot.}
\end{figlist}


\begin{thebibliography}{999}
\def    \nuke   #1#2#3{{\it Nucl. Phys.} {B#1}:#3 (#2)}
\def    \pl     #1#2#3{{\it Phys. Lett.} {#1B}:#3 (#2)}
\def    \prl    #1#2#3{{\it Phys. Rev. Lett.} {#1}:#3 (#2)}
\def    \pr     #1#2#3{{\it Phys. Rev.} {#1}:#3 (#2)}
\def    \prd    #1#2#3{{\it Phys. Rev.} {D#1}:#3 (#2)}
\def    \prep   #1#2#3{{\it Phys. Rep.} {#1}:#3 (#2)}
\def    \zeit   #1#2#3{{\it Z. Phys.} {C#1}:#3 (#2)}
\def    \sovjnp #1#2#3{{\it Sov. J. Nucl. Phys.} {#1}:#3 (#2)}
\def    \cmp    #1#2#3{{\it Comm. Math. Phys.} {#1}:#3 (#2)}
\def    \revm   #1#2#3{{\it Rev. Mod. Phys.} {#1}:#3 (#2)}
\def    \revmp  #1#2#3{{\it Rev. Mod. Phys.} {#1}:#3 (#2)}
\def    \nim    #1#2#3{{\it Nucl. Inst. Meth.} {#1}:#3 (#2)}
\def	\arev	#1#2#3{{\it Annu. Rev. Nucl. Part. Sci.} {#1}:#3 (#2)}
\def	\compc  #1#2#3{{\it Comp. Phys. Comm.} {#1}:#3 (#2)}
\def    \cdf  {(CDF)}
\def    \uaone  {(UA1)}
\def    \uatwo  {(UA2)}
\parskip 0pt
\itemsep=0pt
\small
\baselineskip 12pt
\bibitem{qcd1} Gross~DJ, Wilczek~F, \prl{30}{1974}{1343}
\bibitem{qcd2} Politzer~HD, \prl{30}{1974}{1346}
\bibitem{altarev} Altarelli~G, \arev{39}{1989}{357}
\bibitem{DDT} Dokshitzer~YuL, Dyakonov~DI, Troyan~SI, \prep{58}{1980}{270}
\bibitem{mueller} Mueller~AH, \prep{73}{1981}{237}
\bibitem{altarep} Altarelli~G, \prep{81}{1982}{1}
\bibitem{wilczek} Wilczek~F, \arev{32}{1982}{177}
\bibitem{GLR} Gribov~LA, Levin~EM, Ryskin~MG, \prep{100}{1983}{1}
\bibitem{BCM} Bassetto~A, Ciafaloni~M, Marchesini~G, \prep{100}{1983}{201}
\bibitem{webberrev} Webber~BR, \arev{36}{1986}{253}
\bibitem{collins} Collins~JC, Soper~DE, \arev{37}{1987}{383}
\bibitem{PQCD} Mueller~AH, ed., {\it Perturbative QCD}.
	Singapore: World Scientific (1989), 614 pp.
\bibitem{physrep} Mangano~ML, Parke~SJ, \prep{200}{1991}{301}
\bibitem{leprev} Bethke~S, Pilcher~JE, \arev{42}{1992}{251};
	Hebbeker~T, \prep{217}{1992}{69}
\bibitem{ehlq} Eichten~E, et al., \revm{56}{1984}{579}
\bibitem{dilella} DiLella~L, \arev{35}{1985}{107}
\bibitem{bagnaia} Bagnaia~P, Ellis~SD, \arev{38}{1988}{659}
\bibitem{shapiro} Shapiro~MD, Siegrist~JL, \arev{41}{1991}{97}
\bibitem{altalella} Altarelli~G, DiLella~L, eds.,
	{\it Proton-Antiproton Collider Physics}.
	Singapore: World Scientific (1989)
\bibitem{barger} Barger~V, Phillips~RJN, {\it Collider Physics}. Reading, Mass:
	Addison Wesley (1987)
\bibitem{mishra} Mishra~SR, Sciulli~F, \arev{39}{1989}{259}
\bibitem{tungrev} Tung~WK, Owens~JF, \arev{42}{1992}{291}
\bibitem{ap}
	Altarelli~G, Parisi~G,
	\nuke{126}{1977}{641};
	Dokshitzer~YuL,
	{\it Sov. Phys.} JETP 46:641 (1977);
	Gribov~VN, Lipatov~LN,
	\sovjnp{15}{1972}{438}
\bibitem{giele} Giele~WT, Glover~EWN, Kosower~DA, Fermilab-PUB-92-230-T (1992);
	Giele~WT, Glover~EWN, \prd{46}{1992}{1980}
\bibitem{bernnlo} Bern~Z, Kosower~DA, \nuke{379}{1992}{451};
	\prl{66}{1991}{1669}; \prd{38}{1988}{1888};
	Bern~Z, Dixon~L, Kosower~DA, SLAC-PUB-6001 (1992)
\bibitem{dflm} Diemoz~M, et al., \zeit{39}{1988}{21}
\bibitem{mrs} Martin~A, Roberts~R, Stirling~JW, \prd{37}{1988}{1161}
\bibitem{aurenchepdf} Aurenche~P, et al., \prd{39}{1989}{3275}
\bibitem{grv} Gl\"uck~M, Reya~E, Vogt~A, \zeit{48}{1990}{471}
\bibitem{owens} Owens~J, \pl{266}{1991}{126}
\bibitem{hmrs} Harriman~P, et al., \prd{42}{1990}{798}
\bibitem{kmrs} Kwiecinski~J, et al., \prd{42}{1990}{3645}
\bibitem{mt} Morfin~J, Tung~W, \zeit{52}{1991}{13}
\bibitem{newmrs}
	Martin~A, Roberts~R, Stirling~JW,
	RAL-92-021, DTP/92/16 (1992)
\bibitem{CTEQ} Botts~J, et al., (CTEQ),  MSUHEP-92-27, Fermilab-Pub-92/371
	(1992)
\bibitem{nmc} Amaudruz~P, {et al.}, (NMC),
	CERN Preprint CERN-PPE/92-124.
\bibitem{CCFR} Mishra~SR, et al., (CCFR) Nevis Rep. NEVIS-1466 (1992);
	\zeit{53}{1992}{51}
\bibitem{string} B. Andersson, G. Gustafson and T. Sj\"ostrand,
	\zeit{6}{1980}{235}; \nuke{197}{1982}{45}
\bibitem{webber} Webber~B., \nuke{238}{1984}{492}
\bibitem{LPHD} Azimov~YI, et al., \zeit{27}{1985}{65}; Dokshitzer~YuL,
	Khoze~VA, Troyan~SI, see Ref.~\cite{PQCD}, p.241
\bibitem{amati} Amati~D, Veneziano~G, \pl{83}{1979}{87};
	Marchesini~G, Trentadue~L, Veneziano~G, \nuke{181}{1981}{335}
\bibitem{odorico} Odorico~R, \compc{32}{1984}{139}
\bibitem{earwig} Marchesini~G, Webber~BR, \nuke{238}{1984}{1}
\bibitem{bengtsson} Bengtsson~HU, Ingelman~G, \compc{34}{1985}{251}
\bibitem{isajet} Paige~F, Protopopescu~SD, Brookhaven report BNL-38034 (1986)
\bibitem{field} Field~RD, \nuke{264}{1986}{687}
\bibitem{pythia} Sj\"ostrand~T, Bengtsson~M, \compc{43}{1987}{367}
\bibitem{herwig} Marchesini~G,  Webber~B, \nuke{310}{1988}{461}
\bibitem{CMW} Catani~S, Marchesini~G, Webber~BR, \nuke{349}{1991}{635}
\bibitem{localc} Combridge~BL, Kripfganz~J, Ranft~J, \pl{70}{1977}{234};
	Owens~JF, Reya~F, Gl\"uck~M, \prd{17}{1978}{2324}; \prd{18}{1978}{1501}
\bibitem{ua2jets} Appel~J, {et al.}, \uatwo, \pl{160}{1985}{349};
	Alitti~J, {et al.}, \uatwo, \pl{257}{1991}{232}
\bibitem{ua1jets} Arnison~G, et al., \uaone, \pl{172}{1986}{461}
\bibitem{isrjets} \AA kesson~T, {et al.}, (AFS),
	\pl{123}{1983}{133}
\bibitem{cdfjets} Abe~F, {et al.}, \cdf, \prl{62}{1989}{613}
\bibitem{oldnlo} Ellis~RK, et al, \nuke{173}{1980}{397}
\bibitem{sexton} Ellis~RK, Sexton~J, \nuke{269}{1986}{445}
\bibitem{furman} Sterman~G, Weinberg~S, \prl{39}{1977}{1436};
	Furman~M, \nuke{197}{1982}{413}
\bibitem{aversa} Aversa~F, {et al.}, \pl{210}{1988}{225}
\bibitem{kunzst} Ellis~S, Kunszt~Z, Soper~D, \prl{62}{1989}{2188};
	\prl{64}{1990}{2121}
\bibitem{mjjpaper} Ellis~S, Kunszt~Z,  Soper~D, \prl{69}{1992}{1496}
\bibitem{ua1alg}Arnison~G., {et al.}, \uaone, \pl{132}{1983}{214}
\bibitem{ua2alg} Beer~A, {et al.}, \nim{224}{1984}{360};
	Appel~J, {et al.}, \zeit{30}{1986}{341}
\bibitem{3jetprd} Abe~F, {et al.}, \cdf, \prd{45}{1992}{1448}
\bibitem{huth_snow} Huth~J, {et al.}, in {\it Proc. of the 1990 Summer Study on
    High Energy Physics}, ed E~Berger. Singapore: World Scientific (1992),
p.134
\bibitem{cdfjets} Abe~F, {et al.}, \cdf, \prl{68}{1992}{1104}
\bibitem{wainer} Abe~F, {et al.}, \cdf, Fermilab Report No.
	Fermilab-Pub-92/167-E (1992), submitted to Phys. Rev. Lett.
\bibitem{eksprl} Ellis~S, Kunszt~Z, Soper~D, \prl{69}{1992}{3615}
\bibitem{aversa2}Aversa~F, {et al.}, \zeit{49}{459}{1991}
\bibitem{jadealg} Bartel~W, {et al.}, (JADE),
	\zeit{33}{1986}{23}
\bibitem{durham} Catani~S, et al., \pl{269}{1991}{432};
	Brown~N, Stirling~WJ, \pl{252}{1990}{657} and \zeit{53}{1992}{629};
	Bethke~S, et al., \nuke{370}{1992}{310}
\bibitem{durhamDIS} Catani~S, Dokshitzer~YuL, Webber~BR, \pl{285}{1992}{291}
\bibitem{behrends} Abe~F, {et al.}, \cdf, Fermilab Report No.
        Fermilab-Pub-92/286-E (1992), submitted to Phys. Rev. Lett.
\bibitem{lane} Lane~K, Ramana~M, \prd{44}{1991}{2678}
\bibitem{axigluons} Frampton~P,  Glashow~S, \pl{190}{1987}{157}
\bibitem{orso}Abe~F, {et al.}, \cdf, Fermilab Report No.
        Fermilab-Pub-93/017-E (1993), to be submitted to Phys. Rev. D.
\bibitem{mueller} Abe~F, {et al.}, \cdf, \prl{69}{1992}{2896}
\bibitem{eichten} Eichten~E, {et al.}, \prl{50}{1983}{811}
\bibitem{ua1frag} Bagnaia~P, et al., \uatwo, \pl{144}{1984}{291};
	Arnison~G, et al., \uaone, \nuke{276}{1986}{253}
\bibitem{hubbard} Abe~F, {et al.}, \cdf, \prl{65}{1990}{968}
\bibitem{tasso} Althoff~M, {et al.}, (TASSO),
	\zeit{22}{1984}{307}
\bibitem{doublep} Paver~N, Treleani~D, \zeit{28}{1985}{187};
	Humpert~B, Odorico~R, \pl{154}{1985}{211};
	Landshoff~PV, Polkinghorne~JC, \prd{18}{1978}{3344}
\bibitem{ks} Kunszt~Z, Stirling~W, \pl{171}{1986}{307}
\bibitem{bgk} Berends~F, Giele~WT, Kuijf~H, \nuke{333}{1990}{120};
	\pl{232}{1990}{266}
\bibitem{effpdf} Halzen~F, Hoyer~P,
	\pl{130}{1983}{326};
	Combridge~BL, Maxwell~CJ,
	\nuke{239}{1984}{429}
\bibitem{ptamp}	Parke~S, Taylor~T, \prl{56}{1986}{2459}
\bibitem{ksapp} Kunszt~Z, Stirling~W, \prd{56}{1988}{2439}
\bibitem{maxwell} Maxwell~CJ, \pl{192}{1987}{190}
\bibitem{mpapprox} Maxwell~CJ, \nuke{316}{1989}{321};
	Mangano~ML, Parke~SJ, \prd{39}{1989}{758};
	Maxwell~CJ, Parke~SJ, \prd{44}{1991}{2727}
\bibitem{ua4jets}Alitti~J, {et al.}, \uatwo, \pl{268}{1991}{145}
\bibitem{cdf4jets} Abe~F, {et al.}, \cdf, Fermilab Preprint
	Fermilab-Pub-93/003-E (1993), submitted to Phys. Rev. Lett.
\bibitem{afs4jets} \AA kesson~T, {et al.}, (AFS),
	\zeit{34}{1987}{163}
\bibitem{sjostrand} Sj\"ostrand~T, Sijl~M, \prd{36}{1987}{2019}
\bibitem{sumet} Abe~F, et al., \cdf, \prd{45}{1992}{2249}
\bibitem{dukeowens} Duke~DW, Owens~JF, \prd{30}{1984}{49}
\bibitem{meschi} Meschi~E, \cdf,
	presented at DPF Meet. Am. Phys. Soc., Batavia,
	FNAL-CONF-92/340-E (1992), to appear in the Proceedings.
\bibitem{cpviolation}
	Nir~Y, Quinn~HR,
	\arev{42}{1992}{211}
\bibitem{nason}
	Nason~P,
	in "Heavy Flavours", Advanced Series on Directions in High Energy
	Physics, ed. M~Lindner, AJ~Buras. Singapore: World
       	Scientific (in press)
\bibitem{nellis}
	Ellis~N, Kernan~A,
	\prep{195}{1990}{23}
\bibitem{nde}
	Nason~P, Dawson~S, Ellis~RK,
	\nuke{303}{1988}{607};
	Beenakker~W, et al.,
        \prd{40}{1989}{54}
\bibitem{nde_pt}
	Nason~P, Dawson~S, Ellis~RK,
	\nuke{327}{1988}{49 };
	Beenakker~W, et al.,
        \nuke{351}{1991}{507}
\bibitem{mnr}
	Mangano~M, Nason~P, Ridolfi~G,
	\nuke{373}{1992}{295}
\bibitem{binjets}
	Furmanski~W, Petronzio~R, Pokorski~S,
	\nuke{155}{1979}{253};
	Mueller~AH, Nason~P,
	\nuke{266}{1986}{265};
	Mangano~M, Nason~P,
	\pl{285}{1992}{160}
\bibitem{hvqfrag}
	Azimov~YaI, Dokshitzer~YuL, Khoze~VA,
	\sovjnp{36}{1982}{878}
\bibitem{peterson}
    	Peterson~C, et al.,
	\pr{D27}{83}{105};
	Chrin~J,
	\zeit{36}{1987}{163}
\bibitem{colangelo}
	Colangelo~G, Nason~P,
	\pl{285}{1992}{167}
\bibitem{mele}
	Mele~B, Nason~P,
	\nuke{361}{1991}{626}
\bibitem{kunszt80}
	Kunszt~Z, Pietarinen~E,
	\nuke{164}{1980}{45}
\bibitem{lipatov}
	Lipatov~LN, Frolov~GV,
	\sovjnp{13}{1971}{333};
	Kuraev~EA, Lipatov~LN, Fadin~VS,
	{\it Sov. Phys.} JETP 44:443 (1976); 45:199 (1977);
	Balitskii~YaYa, Lipatov~LN,
	\sovjnp{28}{1978}{822};
	Gribov~LV, Levin~EM, Ryskin~MG,
	\prep{100}{1983}{1};
	Lipatov~LN,
	see Ref.~\cite{PQCD}, p.411
\bibitem{smallx1}
	Ellis~RK, Ross~DA,
	\nuke{345}{1990}{79};
	Collins~JC, Ellis~RK,
	\nuke{360}{1991}{3}
\bibitem{smallx2}
	Catani~S, Ciafaloni~M, Hautmann~F,
	\nuke{366}{1991}{135}
\bibitem{smallx3}
	Levin~EM, Ryskin~MG Shabelsky~YuM,
	\pl{260}{1991}{429}
\bibitem{ua1_b}
	Albajar~C, et al., UA1 Coll.,
	\pl{213}{1988}{405}; \pl{256}{1991}{121}
\bibitem{ua1_Lb}
	Albajar~C, et al., UA1 Coll., \pl{273}{1991}{540}
\bibitem{cdf_psiK}
	Abe~F, et al., (CDF), \prl{68}{1992}{3403}
\bibitem{cdf_psiKstar}
	Vejcik~S, (CDF), presented at DPF Meet. Am. Phys. Soc., Batavia
	(1992), to appear in the Proceedings
\bibitem{cdf_bmu}
	Huffman~BT, (CDF), presented at DPF Meet. Am. Phys. Soc., Batavia,
	FNAL-CONF-92/337-E (1992), to appear in the Proceedings
\bibitem{lqcdLEP}
	Bethke~S, Catani~S,
	presented at {XXVII Rencontres de Moriond}, Les Arcs, CERN-TH-6484-92
	(1992); Altarelli~G, presented at {\it QCD 20 years later}, Aachen,
	CERN-TH.6623/92 (1992)
\bibitem{berger}
	Berger~EL, Meng~R, Tung~WK,
	\prd{46}{1992}{1895};
	Berger~EL, Meng~R, Qiu~J,
	ANL-HEP-CP-92-79.
\bibitem{mlmb}
	Mangano M,
	IFUP-TH-2/93 (1993)
\bibitem{geiser}
	Geiser A,
	presented at XXVII Rencontres de Moriond, Les Arcs, PITHA 92/19 (1992)
\bibitem{onia}
	Guberina~G, et al.,
	\nuke{B174}{1980}{317};
	Berger~EL, Jones~D,
	\prd{23}{1981}{1521};
	Baier~R, R\"uckl~R,
	\zeit{19}{1983}{251};
	Humpert~B, \pl{184}{1987}{105};
	Gastmans~R, Troost~W, Wu~TT,
	\nuke{291}{1987}{731}
\bibitem{gms}
	Glover~EWN, Martin~AD, Stirling~WJ,
	\zeit{38}{1988}{473};
	Glover~EWN, Halzen~F, Martin~AD,
	\pl{185}{1987}{441}
\bibitem{ua1_psi}
	Albajar~C, et al., UA1 Coll.,
	\pl{200}{1988}{380}; \pl{256}{1991}{112}
\bibitem{cdf_psi}
	Abe~F, et al., (CDF), FERMILAB-PUB-92/236-E (1992)
\bibitem{cdf_chi}
	Boswell~C, (CDF), presented at DPF Meet. Am. Phys. Soc., Batavia,
	FNAL-CONF-92/347-E (1992), to appear in the Proceedings
\bibitem{bodwin}
	Bodwin~G, Braaten~E, Lepage~GP,
	\prd{46}{1992}{R1914}
\bibitem{braaten}
	Braaten~E, Yuan~TC,
	NUHEP-TH-92-23; UCD-92-25 (1992)
\bibitem{bchibr}
	R.A. Poling, in {\em Joint International Symposium and Europhysics
	Conference on High Energy Physics}, S Hegarty et al. ed. Singapore:
	World Scientific (1992), p.546.
\bibitem{sps_wxsec}
	Albajar~C, et al., UA1 Coll., \pl{253}{1991}{503};
	Alitti~J, et al., UA2 Coll., \pl{276}{1992}{365}
\bibitem{cdf_wxsec}
	Abe~F, et al., (CDF), \prd{44}{1991}{29};
	\prl{69}{1992}{28}
\bibitem{wxsect}
 	Hamberg~R, Matsuura~T, van Neerven WL,
	\nuke{359}{1991}{343};
	van Neerven~WL, Zijlstra~EB,
	\nuke{382}{1992}{11}
\bibitem{dyresum}
	Sterman~G,
	\nuke{310}{1987}{281};
	Catani~S, Trentadue~L,
	\nuke{327}{1989}{323}
\bibitem{wasy}
	Halzen~F, in Proc. Hadron Structure Functions and Parton Distributions,
        ed. DF~Geesaman et al., World Sci., Singapore (1990);
	Berger~EL, et al.,
	\prd{40}{1989}{83}; Erratum, \prd{40}{1989}{3789}
\bibitem{kevin}
	Einsweiler~K, to appear in this volume.
\bibitem{cdf_wasy}
	Abe~F, et al., (CDF), \prl{68}{1992}{1458}
\bibitem{reno_wasy}
	Baer~H, Reno~MH,
	\prd{43}{1991}{2892}
\bibitem{reno_wpt}
	Arnold~PB, Reno~MH,
	\nuke{319}{1989}{37};
	Gonsalvez~RJ, Paw\-lowski~J, Wai~CF,
	\prd{40}{1989}{2245}
\bibitem{sps_wpt}
	Albajar~C, et al., UA1 Coll., \zeit{44}{1989}{15};
	Alitti~J, et al., UA2 Coll., \zeit{47}{1990}{523}
\bibitem{cdf_wpt}
	Abe~F, et al., (CDF), \prl{66}{1991}{2951}, \prl{67}{1991}{2937}
\bibitem{sudakov}
	Dokshitzer~YuL, Dyakonov~DL, Troyan~SI,
	\pl{78}{1978}{290};
	Parisi~G, Petronzio~R,
	\nuke{154}{1979}{427};
	Kodaira~J, Trentadue~L,
	\pl{112}{1982}{66}
\bibitem{arnold_wpt}
	Arnold~PB, Kaufmann~RP,
	\nuke{349}{1991}{381}
\bibitem{wpt_resum}
	Altarelli~G, et al.,
	\nuke{246}{1984}{12}; \zeit{27}{1985}{617}
\bibitem{cdfwjets} 	Rodrigo~T, \cdf,
	presented at DPF Meet. Am. Phys. Soc., Batavia,
	FNAL-CONF-92/342-E (1992), to appear in the Proceedings.
\bibitem{ua1wjets} Albajar~C, et al., (UA1), \zeit{44}{1989}{15}
\bibitem{dallashuth} Huth~J, (CDF), FERMILAB-CONF-92/326-E (1992), in
	{\it Proc. XXVI Int. Conf. on HEP}, Aug 6-12, 1992, Dallas (TX), to be
	published;
\bibitem{treew}
	Gunion~J, Kunszt~Z, \pl{161}{1985}{333};
	Kleiss~R, Stirling~WJ, \nuke{262}{1985}{235};
	Hagiwara~K, Zeppenfeld~D, \nuke{313}{1989}{560};
	Berends~FA, Giele~WT, Kuijf~H, \nuke{321}{1989}{39};
	Berends~FA, et al. \nuke{357}{1991}{32}
\bibitem{w1jetalpha} Alitti~J, et al., (UA2), \pl{263}{1991}{563}
\bibitem{gammarev}
	Berger~EL, Braaten~E, Field~RD,
	\nuke{239}{1984}{52};
	Ferbel~T, Molzon~WR,
	\revmp{56}{1984}{181};
	Owens~J,
	\revmp{59}{1987}{465}
\bibitem{chapdf}
	Fletcher~RS, Halzen~F, Zas~E,
	\pl{221}{1989}{403}
\bibitem{sps_pho}
	Albajar~C, et al., UA1 Coll.,
	\pl{209}{1988}{385};
	Alitti~J, et al., UA2 Coll.,
	\pl{288}{1992}{386}
\bibitem{cdf_pho}
	Abe~F, et al., (CDF),
	\prl{68}{1992}{2734}; Fermilab-PUB-92/01-E (1992)
\bibitem{inc_pho}
	Aurenche~P, et al.,
	\pl{140}{1983}{87}; \nuke{286}{1987}{509}; \nuke{297}{1988}{661}
\bibitem{iso_pho}
	Aurenche~P, Baier~R, Fontannaz~M,
	\prd{42}{1990}{1440};
\bibitem{berger_iso}
	Berger~EL, Qiu~J,
	\prd{44}{1991}{2002}
\bibitem{pho_zolly}
	Kunszt~Z, Trocsanyi~Z, ETH-TH/92-26 (1992);
	Glover~EWN, Stirling~WJ, \pl{295}{1992}{128}
\bibitem{jet_pho}
	Baer~H, Ohnemus~J, Owens~JF,
	\prd{42}{1990}{61}
\bibitem{pho_nlofrag}
	Aurenche~P, et al.,
	LPTHE-Orsay 92/30, ENSLAPP-A-386/92 (1992)
\bibitem{ua2_pdf}
	Alitti~J, et al., UA2 Coll.,
	CERN-PPE/92-169 (1992)
\bibitem{lhc_dipho}
	Seez~C, et al., in {\it Proc. of LHC Workshop}, eds. G~Jarlskog D~Rein,
	Aachen. CERN 90-10, vol. II p.474 (1990)
\bibitem{cdf_dipho}
	Abe~F, et al., (CDF),
	Fermilab-PUB-92/380-E (1992)
\bibitem{nlo_digamma}
	Aurenche~P, et al.,
	\zeit{29}{1985}{459}
\bibitem{owens_digamma}
	Bailey~B, Ohnemus~J, Owens~JF,
	FSUHEP-920320 (1992)
\end{thebibliography}
\end{document}